\documentclass[12pt]{article}
\usepackage{epsfig,amsmath}
\makeatletter
\renewcommand{\section}{\setcounter{equation}{0}\@startsection
  {section}%
  {1}%
  {0pt}%
  {-1\baselineskip}%
  {0.4\baselineskip}%
  {\bfseries}}%
\renewcommand{\subsection}{\@startsection
  {subsection}%
  {2}%
  {0pt}%
  {-0.75\baselineskip}%
  {0.2\baselineskip}%
  {\bfseries}}%
\renewcommand{\subsubsection}{\@startsection
  {subsubsection}%
  {3}%
  {0pt}%
  {-0.5\baselineskip}%
  {0.1\baselineskip}%
  {\sc}}%
\makeatother

\renewenvironment{itemize}
  {\begin{list}%
     {}%
     {\setlength{\topsep}{-6pt}%
      \setlength{\partopsep}{-6pt}%
      \setlength{\itemsep}{-3pt}%
      \setlength{\labelsep}{5pt}%
      \setlength{\itemindent}{0pt}%
     }%
  }%
  {\end{list}}%

\renewcommand{\theequation}{\thesection.\arabic{equation}}

\def\a{\alpha} 

\def\d{\delta}          

\def\ga{\gamma}         
\def\gm{\Gamma}
 
\def\la{\lambda}        
\def\La{\Lambda}

\def\m{\mu}
\def\n{\nu}

\def\s{\sigma}

\def\tp{\tilde{p}}
\def\tq{\tilde{q}}
\def\hp{\hat{p}}
\def\hm{\hat{M}}
\def\st{\star}

\def\cosutdc{\cos\Big(\frac{p_1\wedge p_3 + p_2\wedge p_4}{2}\Big)}
\def\cosud{\cos\!\left(\frac{p_1\wedge p_2}{2}\right)}
\def\costc{\cos\!\left(\frac{p_3\wedge p_4}{2}\right)}
\def\cosppp{\cos\!\Big(\frac{p_1\wedge p_2 + p_1\wedge p_3 
                              + p_2\wedge p_3}{2}\Big)}
\def\cosppm{\cos\!\Big(\frac{p_1\wedge p_2 + p_1\wedge p_3 
                              - p_2\wedge p_3}{2}\Big)}
\def\cospmm{\cos\!\Big(\frac{p_1\wedge p_2 - p_1\wedge p_3 
                              - p_2\wedge p_3}{2}\Big)}

\def\idk{\int\! \frac{d^4\!k}{(2\pi)^4} \,\,}
\def\idx{\int\! d^4\!x \,}

\def\ds{\displaystyle}

\def\to{\rightarrow}
\def\igual{\hspace{-7pt}=\hspace{-7pt}}
\def\mas{\hspace{-7pt}+\hspace{-7pt}}
\def\menos{\hspace{-7pt}-\hspace{-7pt}}

\def\uvlim{{}_{{}^{\ds \>\longrightarrow}_{\scriptstyle \La\to \infty}}}
\def\PNPo{\bigg\{\!\! \begin{array}{c} 2\\
                 \cos(p\wedge k) \end{array} \!\!\bigg\}}
\def\PNPt{\bigg\{\!\! \begin{array}{c} 2\la_1\\
                (\la_2-\la_1)\,\cos(p\wedge k) \end{array}\!\! \bigg\}}
\def\PNPtr{\bigg\{\!\! \begin{array}{c} 1\\
                 \cos(p\wedge k) \end{array} \!\!\bigg\}}
\def\noverr{\bigg(\begin{array}{c} \!n\!\\\!r\!\end{array}\bigg)}

\begin{document}
\begin{titlepage}
\rightline{UCM-FT/2002-13-01}

\vskip 1.5 true cm
\begin{center}
{\Large \bf UV/IR mixing and the Goldstone theorem\\[9pt] 
in noncommutative field theory}
\vskip 1.2 true cm 
{\rm F. Ruiz Ruiz}
\vskip 0.3 true cm
{\it Departamento de F\'{\i}sica Te\'orica I, 
     Facultad de Ciencias F\'{\i}sicas}\\ 
{\it Universidad Complutense de Madrid, 28040 Madrid, Spain}\\
\vskip 1.2 true cm

{\leftskip=25pt \rightskip=25pt 
\noindent
Noncommutative IR singularities and UV/IR mixing in relation with the
Goldstone theorem for complex scalar field theory are
investigated. The classical model has two coupling constants, $\la_1$
and $\la_2$, associated to the two noncommutative extensions
$\phi^*\!\st\phi\st\phi^*\!\st\phi$ and
$\phi^*\!\st\phi^*\!\st\phi\st\phi$ of the interaction term $|\phi|^4$
on commutative spacetime. It is shown that the symmetric phase is
one-loop renormalizable for all $\la_1$ and $\la_2$ compatible with
perturbation theory, whereas the broken phase is proved to exist at
one loop only if $\la_2=0$, a condition required by the Ward
identities for global $U(1)$ invariance.  Explicit expressions for the
noncommutative IR singularities in the 1PI Green functions of both
phases are given. They show that UV/IR duality does not hold for any
of the phases and that the broken phase is free of quadratic
noncommutative IR singularities. More remarkably, the pion selfenergy
does not have noncommutative IR singularities at all, which proves
essential to formulate the Goldstone theorem at one loop for all
values of the spacetime noncommutativity parameter $\theta$. \par }
\end{center}

\vfil

\end{titlepage}
\setcounter{page}{2}


\section{Introduction}

As is well known, in noncommutative field theory \cite{reviews} the
nonplanar parts of 1PI Green functions become singular when the
noncommutativity spacetime parameter $\theta$ approaches zero
\cite{Minwalla}. The corresponding singularities are called
noncommutative IR divergences and, for the theories usually
considered, are quadratic, linear or logarithmic in $1/\theta$. They
arise from the contribution of large loop-momenta to nonplanar
one-loop Feynman integrals which, being finite for nonvanishing
$\theta$, become divergent if $\theta\to 0$. This simple but deep
observation, first made in ref. \cite{Minwalla}, is known as UV/IR
mixing and for $\la\phi^4$ and gauge theories
\cite{Minwalla}-\cite{vac} takes a much stronger form, which we will
refer to as strong UV/IR duality. Strong UV/IR duality states that the
logarithmic noncommutative IR singularities in the nonplanar part of a
1PI Green function and the logarithmic UV divergences in its planar
part are in one-to-one correspondence. UV/IR duality in this strong
form seems not to be an artifact of perturbation theory, since in many
instances it has been reobtained by taking the infinite tension limit
of a suitable string amplitude for an open bosonic string on a
magnetic $B$-field \cite{Andreev}.

Noncommutative IR singularities pose serious problems for the
existence of noncommutative field theories beyond one loop. They
threaten renormalizability at higher loops (since locality of UV
counterterms may be spoiled\footnote{As of today, the question of
higher-loop renormalizability has been addressed mainly for $\la\phi^4$
\cite{2loop} and the Wess-Zumino model \cite{Girotti}.}) and may
introduce tachyonic states \cite{Matusis} \cite{vac} \cite{Mehen}
(associated to quadratic noncommutative IR singularities in 1PI
two-point functions). In noncommutative gauge theories \cite{gauge},
quadratic and linear noncommutative IR singularities can be eliminated
by introducing supersymmetry \cite{Matusis} \cite{vac}. Indeed, in
supersymmetric gauge theories, the supersymmetric partners of the
gauge field provide nonplanar contributions which cancel the quadratic
and linear noncommutative IR singularities in the nonsupersymmetric
theories \cite{Matusis} \cite{vac}. The supersymmetric theories thus
become free of tachyonic instabilities and are left with the milder
noncommutative logarithmic IR singularities.  Furthermore, the results
in ref. \cite{vac} imply that supersymmetric $N=1$ $U(1)$ gauge theory
in the Yennie gauge becomes free of {\it all} noncommutative IR
singularities at one loop.

The purpose of this paper is to study the noncommutative IR
singularities of $U(1)$ complex scalar field theory, to investigate
whether they satisfy UV/IR duality in the strong sense mentioned
above, to explore spontaneous symmetry breaking as a mechanism to
eliminate noncommutative IR singularities and to analyze how this
enters the Goldstone theorem. To carry this investigation, we must
first understand the UV renormalizability of the model. Although the
latter should be by now well established, in our analysis we
have found issues that have gone unnoticed in the literature and that
are essential to understand the model's spontaneous symmetry breaking
and its $U(1)$ global invariance at the quantum level. We also report
on them.

To be more explicit, consider complex scalar field theory on
noncommutative Minkowski spacetime, defined classically by the action
\begin{equation}
   S_{\rm sym} = \idx \left[ \, 
         (\partial_\m \phi^*)\,( \partial^\m\!\phi) 
                  - V_{\rm sym}(M,\la,\phi,\phi^*)\,\right] ~,
\label{sym-action}
\end{equation}
where $\phi$ is a complex scalar field and the potential
$V_{\rm sym}(M,\la_1,\la_2,\phi,\phi^*)$ has the form
\begin{equation}
   V_{\rm sym}(M,\la_1,\la_2,\phi,\phi^*) = M^2\,|\phi|^2  
     + \frac{\la_1}{4} \> \phi^*\!\st\phi\st\phi^*\!\st\phi
     + \frac{\la_2}{4} \>\phi^*\!\st\phi^*\!\st\phi\st\phi ~,
\label{sym-pot}
\end{equation}
with $\la_1$ and $\la_2$ two different coupling constants. Note that
in the action one must allow for the two inequivalent noncommutative
extensions $\phi^*\!\st\phi\st\phi^*\!\st\phi$ and $\phi^*\!\st\phi^*
\!\st\phi\st\phi$ of the commutative interaction term $|\phi|^4$. The
symbol $\!\st\!$ denotes the Moyal product, defined for functions
$f(x)$ and $g(x)$ as
\begin{displaymath}
  \big(f\!\st\! g\big) (x) = f(x)\, \exp\Big( \frac{i}{2}~
    \overleftarrow {\partial_\m} \, \theta^{\m\n}\,
    \overrightarrow{\partial_\n}\Big) \, g(x)~,
\end{displaymath}
where $\theta^{\m\n}$ is a constant real antisymmetric matrix and our
metric convention is $g_{\m\n}={\rm diag}\>(+,-,-,-)$. We will
restrict ourselves to magnetic-like matrices\footnote{In this way we
do not run into problems with unitarity \cite{Gomis}.}
$\theta^{\m\n}$, {\it i.e.} such that $\theta^{0i}=0$ for
$i=1,2,3$. For $M^2>0$, the only field configuration that minimizes
the energy is $\phi_0=0$ and the action (\ref{sym-action}) with
potential (\ref{sym-pot}) defines the symmetric phase of the classical
theory. The global $U(1)$ gauge transformations that leave invariant
the action take the form $\phi \to e^{i\a}\phi$, with $\a$ an
arbitrary real constant. By contrast, for $M^2<0$, any field
configuration $\phi_0$ such that $\,|\phi_0|^2 = v^2\!$, with
\begin{displaymath}
    v= \sqrt{\frac{-2M^2}{\la_1 + \la_2}} ~,
\end{displaymath}
minimizes the energy and classical spontaneous symmetry breaking takes
place. Indeed, choosing $\phi_0=v$ and expanding $\phi$ about it as
\begin{equation}
   \phi = \frac{1}{\sqrt{2}}\> (\pi+i\s) + iv ~, 
\label{br-fields}
\end{equation}
the action can be written as
\begin{equation}
    S_{\rm br} = \idx \left[ 
          \frac{1}{2}\> (\partial_\m \pi)\, (\partial^\m \pi)
        + \frac{1}{2}\> (\partial_\m \s)\, (\partial^\m \s)
        - V_{\rm br}(M,\la_1,\la_2,\pi,\s) \right] ~,
\label{br-action}
\end{equation}
where the potential $V_{\rm br}(M,\la_1,\la_2,\pi,\s)$ has the form
\begin{equation}
\begin{array}{l}
    {\ds V_{\rm br}(M,\la_1,\la_2,\pi,\s) = \frac{1}{2}\> (2M^2)\,\s^2 
      + \frac{v\,(\la_1+\la_2)}{2\sqrt{2}}\> \big( 
             \pi\!\st\!\pi\!\st\!\s + \s\!\st\!\s\!\st\!\s \big) }\\[12pt]
\hphantom{ V_{\rm br}(M,\la_1,\la_2,\pi,\s)~}
    {\ds +\> \frac{\la_1}{4}\> \pi\!\st\!\pi\!\st\! \s\!\st\!\s 
         - \frac{\la_1-\la_2}{8}\> \pi\!\st\!\s\!\st\!\pi\!\st\!\s }\\[12pt]
\hphantom{ V_{\rm br}(M,\la_1,\la_2,\pi,\s)~}
    {\ds +\> \frac{\la_1+\la_2}{16}\> \big( 
                  \pi\!\st\!\pi\!\st\!\pi\!\st\!\pi 
                + \s\!\st\!\s\!\st\!\s\!\st\!\s \big) }  
\end{array}
\label{br-pot}
\end{equation}
and $M^2$ has been replaced with $-M^2\!$, so as to work with a
positive $M^2$. The action (\ref{br-action}) with potential
(\ref{br-pot}) defines the nonsymmetric or broken phase of the
classical theory. The global $U(1)$transformations that leave $S_{\rm
br}$ invariant are obtained from $\phi \to e^{i\a}\phi$ and
eq. (\ref{br-fields}); they read
\begin{equation}
  \d\pi = -\a\,(\s+\sqrt{2}\,v) \qquad \d\s = \a\pi ~.
\label{br-transformation}
\end{equation}
As stated, we want to study the noncommutative IR singularities and
their mixing with UV divergences in both phases.

Our main results and the organization of the paper are as follows. In
section 2, we consider the symmetric phase and show that it is
one-loop renormalizable for arbitrary $\la_1$ and $\la_2$ compatible
with perturbation theory, being not necessary to take $\la_2=0$. We
also give explicit expressions for the noncommutative IR singularities
in the 1PI Green functions and prove that UV/IR duality in its strong
form does not hold. Sections 3 to 5 are dedicated to study the broken
phase. In particular, in section 3, we demonstrate that one-loop UV
renormalization for the broken phase is consistent with the Ward
identities only if $\la_2=0$. In section 4 we rederive the same result
by analyzing the consistency of the nonplanar sector of the theory
with the Ward identities. Section 5 presents explicit expressions for
the noncommutative IR singularities in the 1PI Green functions of the
broken phase. The expressions given there show that in the the broken
phase there are no quadratic noncommutative IR singularities, that the
selfenergy for the pion field $\pi$ is free of {\it all}
noncommutative IR singularities and that the strong version of UV/IR
duality does not hold. Also in section 5 we show that the pion mass,
defined as the zero of the selfenergy, remains zero after one-loop
radiative corrections, thus ensuring that the Goldstone theorem holds
true at one loop for arbitrary magnetic $\theta^{\m\n}$. Section 6
contains our conclusions.

Several related problems have been addressed in the
literature. 
%
%
In ref. \cite{Campbell} the broken phase of the noncommutative global
$U(N)$ model, with $N>1$ and $\la_2=0$, is considered and it is shown
that the pion selfenergy vanishes for vanishing external
momentum. Ref. \cite{Sarkar} assumes $\la_2=0$ and proves that global
$O(2)$ scalar field theory is one-loop renormalizable. Whereas these
papers deal with the case $\la_2=0$, we focus on the case $\la_2\neq
0$ and on noncommutative IR singularities and their implications.  As
concerns local models, ref. \cite{Petriello} proves the consistency of
UV renormalization with the BRS identities for the local $U(1)$ model
and calculates the beta functions. In turn, the one-loop
renormalizability of the local $U(2)$ and $U(1)\!\times\!  U(1)$
models is shown in ref. \cite{Liao}. It is worth noting that in the
local models $\la_2$ is excluded classically, since
$\phi^*\!\st\phi^*\!\st\phi\st\phi$ is not invariant under local gauge
transformations, while our analysis here shows that in the global
model $\la_2=0$ follows from the symmetry requirements at the quantum
level.

\section{The symmetric phase: renormalization and noncommutative IR
singularities}

We first consider the symmetric phase, with classical action given by
eq. (\ref{sym-action}) and (\ref{sym-pot}). At one loop, the only 1PI
Green functions with UV divergences are the field selfenergy
$\Sigma(p)$ and the vertex $\gm(p_1,p_2,p_3,p_4)$. To regularize the
theory and to account for the counterterms that will be necessary to
subtract the UV divergences, we introduce an invariant cutoff $\La$ by
considering the `bare' action
\begin{equation}
   S_{\La,0}
     = \idx \left[ \, (\partial_\m \phi^*_0) 
       \left( 1+\frac{\partial^2}{\La^2}\right)^n ( \partial^\m\!\phi_0) 
     - V_{\rm sym}(M_0,\la_{10},\la_{20},\phi_0,\phi_0^*)\,\right] 
   \qquad n\geq 2~.
\label{sym-bare-action}
\end{equation}
Note that the quadratic UV divergences in the one-loop tadpole are not
regularized if $n=1$, so we must take $n\geq 2$.  The potential
$V_{\rm sym}(M_0, \la_{1,0}, \la_{2,0}, \phi_0)$ is as in
eq. (\ref{sym-pot}) but with the renormalized quantities
$M,\,\la_1,\,\la_2,\,\phi,\,\phi^*$ replaced with bare quantities
$M_0,\,\la_{10},\,\la_{20},\,\phi_0,\,\phi^*_0$, defined by
\begin{eqnarray}
   & \phi_0 = Z_\phi^{1/2} \,\phi  & \label{sym-bare-phi}\\
   &  Z_\phi M_0^2 = Z_{M^2} M^2 \qquad 
      Z_\phi^2 \la_{10} = \la_1 + \d\la_1 \qquad
      Z_\phi^2 \la_{20} = \la_2 + \d\la_2 ~. & \label{sym-bare-param}
\end{eqnarray}
The renormalization constants $ Z_\phi$ and $ Z_{M^2}$ have the form 
\begin{displaymath}
  Z_\phi = 1 + \d z_\phi \qquad  Z_{M^2} = 1 + \frac{\d M^2}{M^2} ~,
\end{displaymath}
with $\d\la_1,\,\d\la_2,\,\d z_\phi$ and $\d M^2$ collecting all terms of
order one or higher in $\hbar$. The action $S_{\La,0}$ can be
recast as
\begin{displaymath}
   S_{\La,0} =  S_{\La,\rm sym} + S_{\rm ct,sym} ~,
\end{displaymath}
where $S_{\La,\rm sym}$ is given by  
\begin{equation}
    S_{\La,\rm sym} =  \idx \left[ \, (\partial_\m \phi^*) 
       \left( 1+\frac{\partial^2}{\La^2}\right)^n ( \partial^\m\!\phi) 
     - V_{\rm sym}(M,\la_1,\la_2,\phi,\phi^*)\,\right] 
\label{sym-reg-action}
\end{equation}
and the counterterms $S_{\rm ct,sym}$ read
\begin{equation}
  S_{\rm ct,sym} =  \idx \bigg[ 
      \d z_\phi \, (\partial_\m \phi^*) (\partial^\m\!\phi) 
    - \d M^2\, \phi^*\phi 
    - \frac{\d\la_1}{4}\> \phi^*\!\st\phi\st\phi^*\!\st\phi 
    -  \frac{\d \la_2}{4}\> 
                \phi^*\!\st\phi^*\!\st\phi\st\phi \bigg] \>. 
\label{sym-ct-action}
\end{equation}
It is important to emphasize that $\la_1$ and $\la_2$ are different
coupling constants, so there is no reason for them to have the same
running. In other words, $\d\la_1$ and $\d\la_2$ may be
different. Eqs. (\ref{sym-reg-action}) and (\ref{sym-ct-action})
provide the Feynman rules depicted in fig. 1, where we have
used the notation
\begin{displaymath} 
   \tp = \theta^{\m\n}p_\n \qquad p\wedge q = \theta^{\m\n} p_\m q_\n 
   \qquad 
   p\circ p =-\,\theta^{\m\n}\,\theta_{\m}^{~\,\tau}\,p_\n\,p_\tau~.
\end{displaymath}

Introducing sources  $J_0$ and $J_0^*$ for the fields
$\phi_0^*$ and $\phi_0$, we consider the generating functional
\begin{equation}
   Z[J_0,J_0^*] = e^{G_c[J_0,J_0^*]} = \int [d\phi_0]\>
      [d\phi_0^*] \> \exp\left[ iS_{\La,\rm sym} +i S_{\rm ct,sym} 
           + i \idx (J_0^*\phi_0 +J_0\phi_0^*)\,\right] \>.
\label{generating}
\end{equation}
For $J_0$ we write $J_0=Z^{-1/2}_\phi J$, so that $J^*_0\phi_0 =
J^*\phi$, and similarly for $J^*_0$. To find the Ward identity
associated to the $U(1)$ global symmetry, we follow the standard
procedure: change variables $\phi\to e^{i\a}\phi$ in the integral in
eq. (\ref{generating}), take into account that under this change
$S_{\La,0}$ remains invariant and define the effective action
$\gm[\phi,\phi^*]$ as the Legendre transform of $W[J,J^*]$. This leads
to the Ward identity
\begin{equation}
    \idx \left( \phi\>\frac{\d\gm}{\d\phi} 
              - \phi^*\>\frac{\d\gm}{\d\phi^*} \right) =0~.
\label{sym-WI}
\end{equation}
Using for the effective action its expansion 
\begin{displaymath}
\begin{array}{l}
  {\ds \gm[\phi,\phi^*] = \sum_{n=1}^\infty \frac{1}{2n!} 
    \int d^4x_1\cdots d^4x_n\, d^4y_1\cdots d^4y_n\> }\\[12pt]
\hphantom{ \gm[\phi,\phi^*]\,}
    \times \phi(x_1)\cdots \phi(x_n)\> 
    \phi^*(y_1)\cdots \phi^*(y_m) \>
    \gm^{(n,m)}(x_1,\ldots,x_n;y_1,\ldots,y_m) 
\end{array}
\end{displaymath}
in fields, where $\gm^{(n)}(x_1,\ldots,x_n;y_1,\ldots,y_n)$ denotes
the Green function of $n$ $\phi$-fields and $n$ $\phi^*\!$-fields, and
going to momentum space, we obtain the following set of Ward
identities for the 1PI Green functions:
\begin{equation}
  \gm^{(n)}(p_1,\ldots,p_n;q_1,\ldots,q_n) 
     =  \gm^{(n)}(q_1,\ldots,q_n;p_1,\ldots,p_n)~.
\label{WI-1pi-sym}
\end{equation}
The quantum theory is defined by the $\La\to\infty$ limit of
$Z[J_0,J^*_0]$, or equivalently of $\gm[\phi,\phi^*]$. Hence, for the
symmetric phase of the quantum theory to exist, the large $\La$ limit
must be well defined. This means that, while preserving the Ward
identities, it must be possible to choose order by order in
perturbation theory the counterterms so as to cancel the divergences
that appear in the 1PI Green functions when $\La\to\infty$.  We are
going to show that this is the case at one loop for all values of
$\la_1$ and $\la_2$ compatible with perturbation theory.

As already mentioned, the only 1PI Green functions with UV divergences
at one loop are the field selfenergy $\Sigma(p)$ and the vertex
$\gm(p_1,p_2;p_3,p_4)$. Let us first worry about the selfenergy. Its
one-loop contribution is given by
\begin{equation}
   \epsfig{file=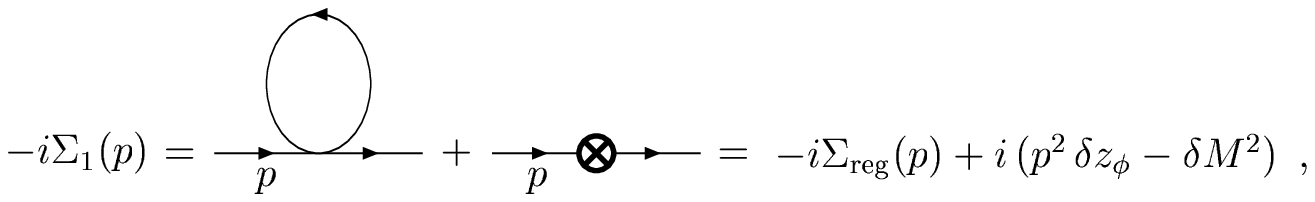} 
\label{2-phi-correction}
\end{equation}
where the regularized selfenergy $-i\Sigma_{\rm reg}(p)$ reads
\begin{displaymath}
   -i\Sigma_{\rm reg}(p) = \idk \frac{\la_1 + {\ds \frac{\la_2}{2}}\>
                             \Big(1+e^{ik\wedge p}\Big)}
                  {k^2\Big(1-{\ds \frac{k^2}{\La^2}}\Big)^n - M^2}  ~.
\end{displaymath}
The $\theta^{\m\n}\!$-independent part of this integral gives the
one-loop planar contribution $-i\Sigma_{\rm P}(p)$ to the field
selfenergy, while the $\theta^{\m\n}\!$-dependent part defines the
nonplanar contribution $-i\Sigma_{\rm NP}(p)$.  Computing their limit
$\La\to\infty$ (see the Appendix for details), we have
\begin{equation}
\begin{array}{l}
  {\ds -i\Sigma_{\rm P}(p) = \Big( \la_1 + {\ds \frac{\la_2}{2}} \Big)  
     \idk \frac{1}{k^2\Big(1-{\ds \frac{k^2}{\La^2}}\Big)^n - M^2} } 
                                                         \\[21pt]
\hphantom{ -i\Sigma_{\rm P}~~~}
      {\ds \uvlim  - \frac{i}{16\pi^2} \Big(\la_1 
         +\frac{\la_2}{2}\Big) \left[ \frac{\La^2}{n-1} 
         - M^2 \ln\Big(\frac{\La^2}{M^2}\Big) + M^2 f_0 \right] } 
\end{array}
\label{sym-2-phi-uv}
\end{equation}
and
\begin{equation}
  -i\Sigma_{\rm NP}(p) = \frac{\la_2}{2}  \idk \frac{e^{i k\wedge p}}
      {k^2\Big(1-{\ds \frac{k^2}{\La^2}}\Big)^n - M^2}  ~~\uvlim ~~  
      -\frac{i\la_2 M^2}{8\pi^2}      
      \frac{K_1(\sqrt{p\circ p\,M^2})}{\sqrt{p\circ p\,M^2}} ~,
\label{sym-2-phi-np} 
\end{equation}
where 
\begin{equation}
   f_0 = \sum_{r=1}^{n-1} \frac{1}{r} + \sum_{r=1}^n \,
      \noverr \, \frac{\gm(r)\,\gm(2n-r)}{\gm(2n)}  
\label{f0}
\end{equation}
and $K_\n(\cdot)$ is the third Bessel function of order $\n$. Note
that, when $\La\to\infty$, the planar contribution diverges
quadratically and the nonplanar contribution remains finite provided
$\,p\circ p \neq 0$. To cancel the UV divergences in $-i\Sigma_{\rm
P}(p)$ and thus render $-i\Sigma_1(p)$ UV finite, we adopt an MS type
scheme and take for $\d z_\phi$ and $\d M^2$
\begin{equation}
  \d z_\phi =0 \qquad \quad
  \d M^2 = -\frac{1}{16\pi^2}\Big( \la_1 +\frac{\la_2}{2}\Big)\, 
       \left[ \frac{\La^2}{n-1} 
            - M^2 \ln\Big(\frac{\La^2}{M^2}\Big) \right] ~.
\label{deltas-2}
\end{equation}

For the one-loop correction to the 4-vertex
\begin{displaymath}
   \epsfig{file=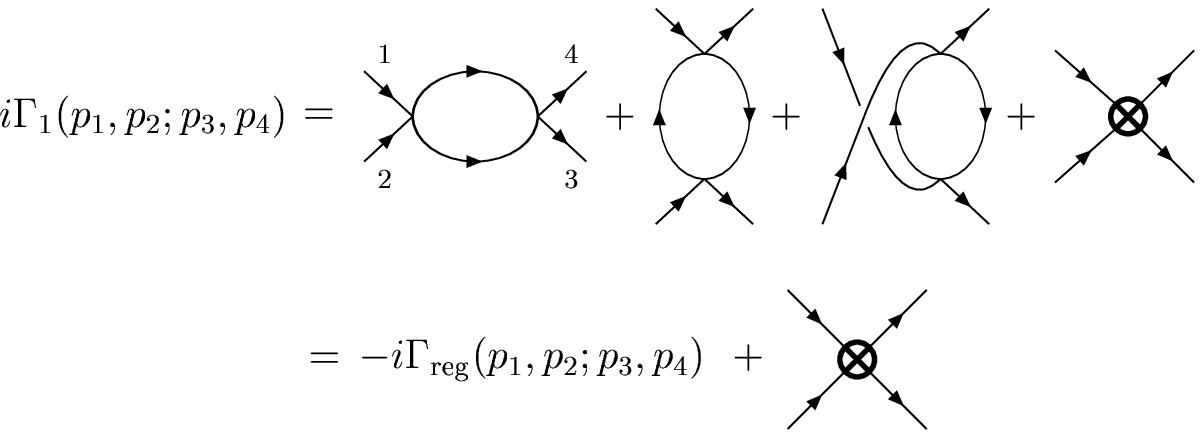} 
\end{displaymath}
we proceed similarly. The regularized contribution
$-i\gm_{\rm reg}(p_1,p_2,p_3,p_4)$ is the sum of the first three
diagrams and can be decomposed 
\begin{equation}
    - i\gm_{\rm reg}(p_1,p_2;p_3,p_4) =- i\gm_{\rm P}(p_1,p_2;p_3,p_4)
                             - i\gm_{\rm NP}(p_1,p_2;p_3,p_4)
\label{sym-4-phi}
\end{equation}
in a planar part $-i\gm_{\rm P}(p_1,p_2;p_3,p_4)$ and a nonplanar part
$-i\gm_{\rm NP}(p_1,p_2;p_3,p_4)$. The planar contribution contains
all the divergences that arise when $\La\to\infty$, while the
nonplanar contribution is well defined for $\La\to\infty$ and
$\theta^{\m\n}\!\neq 0$. After some calculations, for the planar
contribution we obtain
\begin{equation}
\begin{array}{l}
   {\ds -i\gm_{\rm P}(p_1,p_2;p_3,p_4) ~\uvlim~ \frac{i}{16\pi^2}\> 
     \bigg[ \Big( \la_1^2 + \frac{\la_2^2}{4} \Big) \cosutdc }\\[12pt]
\hphantom{-i\gm_{\rm P}(p_1,p_2,p_3,p_4)~~~}
   {\ds +\, \la_2\,\Big(\la_1+\frac{\la_2}{4}\Big)\, \cosud \costc 
         \bigg] \ln\Big(\frac{\La^2}{M^2}\Big) + {\rm f.c.}}~,
\end{array}
\label{sym-4-phi-p}
\end{equation}
where ``f.c.'' collects finite, regular contributions for
nonexceptional configurations of external momenta. In the MS type
scheme that we have adopted, cancellation of UV divergences requires
taking $\d\la_1$ and $\d\la_2$ as
\begin{equation}
\begin{array}{l}
  {\ds \d\la_1 = \frac{1}{16\pi^2}\> 
     \Big( \la_1^2 + \frac{\la_2^2}{4} \Big)\,
            \ln\Big(\frac{\La^2}{M^2}\Big)}\\[12pt]
  {\ds \d\la_2 = \frac{1}{16\pi^2}\> 
     \la_2\,\Big(\la_1+\frac{\la_2}{4}\Big)\,
            \ln\Big(\frac{\La^2}{M^2}\Big)} ~.
\end{array}
\label{deltas-lambdas}
\end{equation}

Thus far we have that the counterterms $S_{\rm ct,sym}$ with $\d
z_\phi$, $\d M^2$, $\d\la_1$ and $\d\la_2$ as in eqs. (\ref{deltas-2})
and (\ref{deltas-lambdas}) cancel the divergences that occur in the
one-loop 1PI diagrams generated by $S_{\La,\rm sym}$ when
$\La\to\infty$, thus ensuring that the $\La\to\infty$ limit of
$\gm[\phi,\phi^*]$ exists at one loop. Furthermore, since by
construction $\gm[\phi,\phi^*]$ satisfies the Ward identity
(\ref{sym-WI}) for all $\La,\,\la_1$ and $\la_2$, and since the
divergent contributions for $\La\to\infty$ in the 1PI Green functions,
given by eqs. (\ref{sym-2-phi-uv}) and (\ref{sym-4-phi-p}), satisfy
the Ward identities (\ref{WI-1pi-sym}), the limit $\La\to\infty$
preserves the Ward identities. Hence the symmetric phase of the
quantum theory exists at one loop. We stress that the symmetric phase
is renormalizable at one loop for all values of $\la_1$ and $\la_2$
compatible with perturbation theory, and that there is no need to
assume $\la_2=0$. In other words, if one writes
\begin{displaymath}
\begin{array}{l}
   \la_1 +\d\la_1 = Z_{11}\la_1 + Z_{12}\la_2 \\
   \la_2 +\d\la_2 = Z_{21}\la_1 + Z_{22}\la_2 ~,
\end{array}
\end{displaymath}
the fact that $\la_1$ and $\la_2$ are different coupling constants
means that there are no requisites on the $Z_{ij}$ other than those
arising from the Ward identities, and we have seen that these do not
impose any. From eqs. (\ref{deltas-lambdas}) we obtain
\begin{displaymath}
\begin{array}{ll}
  {\ds Z_{11} = 1 + \frac{1}{16\pi^2}\>\la_1\> 
                    \ln\Big(\frac{\La^2}{M^2}\Big)} \qquad &
  {\ds Z_{12} = \frac{1}{16\pi^2} \> \frac{\la_2}{4}\>
                \ln\Big(\frac{\La^2}{M^2}\Big) }\\[12pt]
  {\ds Z_{21} = \frac{1}{16\pi^2} \> \frac{\la_2}{2} \>
                 \ln\Big(\frac{\La^2}{M^2}\Big) } &
  {\ds Z_{22} = 1 + \frac{1}{16\pi^2} \> \frac{2\la_1+ \la_2}{4}\> 
                    \ln\Big(\frac{\La^2}{M^2}\Big)~, }
\end{array}
\end{displaymath}
which are different among themselves. In sections 3 and 4, we will see
that for the broken phase the Ward identities require $\la_2=0$.

Once we know that the symmetric phase of the theory exists at one
loop, we move on to study the noncommutative IR singularities in the
1PI Green functions. The one-loop nonplanar contribution
(\ref{sym-2-phi-np}) to the selfenergy is well defined for
$\theta^{\m\n}\!\neq 0$. For $\theta^{\m\n}\to 0$, however, it becomes
singular. In fact, sending $\theta^{\m\n}\to 0$ in eq.
(\ref{sym-2-phi-np}) and using the results in the Appendix, we have
\begin{equation}
  \lim_{\theta^{\m\n}\to 0}\>\lim_{\La\to\infty} \Big[ 
   -i\Sigma_{\rm NP}(\theta,p) \Big] = - 
       \frac{i\la_2}{8\pi^2}\> \left\{ \frac{1}{p\circ p} 
          + \frac{M^2}{4}\>\Big[ \ln(p\circ p\, M^2) 
               -2\ln2 +2\ga - 1 \Big] \right\} ~.
\label{sym-2-phi-ir}
\end{equation}
The origin of these noncommutative IR singularities can be understood
by looking at the integral expression for $-i\Sigma^{\rm NP}(p)$ in
eq. (\ref{sym-2-phi-np}). At $\La\to\infty$, the integral is well
defined if $\theta^{\m\n}\neq 0$, but diverges quadratically if
$\theta^{\m\n}= 0$. The contribution to the integral from arbitrarily
high momenta $k^\m$ is curbed by the noncommutativity of spacetime,
with $1/p\circ p$ acting as a regulator. This is precisely the UV/IR
mixing argument \cite{Minwalla}, that for $\la\phi^4$ and gauge
theories \cite{Minwalla}-\cite{vac} goes beyond this observation for
nonplanar integrals and states that the logarithmic noncommutative IR
singularities in the nonplanar part of a 1PI Green function and the
logarithmic UV divergences in its planar part can be obtained from
each other by replacing $p_i\circ p_i\leftrightarrow 1/\La^2$ for all
the external momenta $p_i$. This stronger form of UV/IR mixing does
not hold here, since the planar part $-i\Sigma_{\rm P}(p)$ of the
selfenergy has UV logarithmic divergences proportional to $\la_1$,
whereas the nonplanar part $-i\Sigma_{\rm NP}(p)$ does not have
contributions proportional to $\la_1$ [see eqs. (\ref{sym-2-phi-uv})
and (\ref{sym-2-phi-np})].
%
%
Without loss of generality, we can take a reference frame in which all
the components of $\theta^{\m\n}$ vanish except for
\begin{equation}
    \theta^{12}=-\theta^{21}\equiv \theta.
\label{frame}
\end{equation} 
In this frame, and using the notation $p^\m=(p^0,\vec{p}_\bot,p^3)$
and $\vec{p}_\bot=(p^1,p^2)$, eq. (\ref{sym-2-phi-ir}) takes the form
\begin{equation*}
 -i\Sigma_{\rm NP}(p)\approx -
    \frac{i\la_2}{8\pi^2}\> \left[ 
        \frac{1}{\theta^2 \vec{p}_\bot^2}
          + \frac{M^2}{2}\> \ln(\theta M^2) \right] ~,
\end{equation*}
where the symbol $\approx$ denotes that the limit $\theta^{\m\n}\to
0,\,\La\to\infty$ has been taken and that all finite contributions
have been dropped. Besides the selfenergy, the four-vertex is the only
other 1PI Green function that may develop noncommutative IR
singularities in its nonplanar part. After some calculus, for the
singular behaviour at $\theta\to 0$ of its nonplanar part, we obtain
in the frame (\ref{frame})
\begin{equation}
\begin{array}{l}
  {\ds -i\gm_{\rm NP}(p_1,p_2,p_3,p_4) \approx -\, \frac{i}{16\pi^2} \>
    \bigg[ \, \la_2 \Big(\la_1+ \frac{3}{8}\,\la_2\Big)\, \cosutdc  
  } \\[12pt]
\phantom{-i\gm_{\rm NP}(p_1,p_2,p_3,p_4)\approx \>}
  {\ds +\, \Big(\,\frac{3}{4}\,\la^2_1 + \la_1\la_2 
     + \frac{5}{8}\,\la_2^2 \Big) \, \cosud \costc\!\! \bigg] 
     \ln\,(\theta M^2) }\>.
\end{array}
\label{sym-4-phi-np}
\end{equation}
It is clear that the replacement $\theta^2M^2\leftrightarrow 1/\La^2$
does not relate the noncommutative IR singularities in this equation
with the UV divergences in the planar part given in
eq. (\ref{sym-4-phi-p}). We conclude that UV/IR duality in its strong
form does not hold.

To finish our discussion of noncommutative IR divergences, we study if
these introduce perturbative tachyonic instabilities as in
nonsupersymmetric gauge theories. The dispersion relation up to one
loop reads
\begin{equation*}
p^2 - M^2 - \Sigma_1(p) =0 ~.
\end{equation*}
For external momenta $p^\m$ such that $\la_2/p\circ p\, M^2 <\!\!< 1$,
where perturbation theory is valid, the dominant part of $\Sigma_1(p)$
is the first term in eq. (\ref{sym-2-phi-ir}), so we write
\begin{displaymath}
   p^2 =   M^2 + \frac{\la_2}{8\pi^2\,p\circ p} 
       + \,{\rm subleading~terms}~.
\end{displaymath}
Since $p\circ p$ is positive definite, there are no perturbative
tachyonic instabilities.

\section{The broken phase: UV counterterms}

We start writing an action analogous to $S_{\La,0}$ for the symmetric
phase which (i) generates through perturbation theory finite Green
functions at $\La\to\infty$, and (ii) is symmetric under global
$U(1)$ transformations. To this end, we combine
eqs. (\ref{sym-bare-phi}) and (\ref{br-fields}) so that
\begin{equation}
   \phi_0 = Z_\phi^{1/2}\, \left[ \frac{1}{\sqrt{2}}\>
           (\pi+i\s) + i v \right] ~, 
\label{Zv}
\end{equation}
substitute the latter in eq. (\ref{sym-action}), use eq.
(\ref{sym-bare-param}) and replace $M^2$ with $-M^2$. This yields for
$S_{\La,0}$
\begin{displaymath}
  S_{\La,0} = S_{\La,\rm br} +  S_{\rm ct,br} ~,
\end{displaymath}
where $S_{\La,\rm br}$ is given by
\begin{equation}
   S_{\La,\rm br} =  \idx \left[ \frac{1}{2}\> (\partial_\m\pi)
        \Big( 1+ \frac{\partial^2}{\La^2}\Big)^n (\partial^\m\pi)
      +\frac{1}{2}\> (\partial_\m\s) \Big( 1 
        + \frac{\partial^2}{\La^2}\Big)^n (\partial^\m\s)
        - V_{\rm br}(M,\la_1,\la_2,\pi,\s) \right] ~,
\label{br-reg-action}
\end{equation}
the counterterms $S_{\rm ct,br}$ read 
\begin{equation}
\begin{array}{l}
  {\ds  S_{\rm ct,br} = \idx \bigg\{ 
     \frac{\d z_\phi}{2} \> (\partial_\m\pi)\, (\partial^\m\pi)
   + \frac{\d z_\phi}{2} \> (\partial_\m\s)\, (\partial^\m\s)
   - \sqrt{2}v\d_1\,\s - \frac{\d_1}{2}\>\pi^2
   - \frac{\d_2}{2}\>\s^2 }\\[12pt]
\hphantom{S_{\rm ct,br} ~}
  {\ds -\, \frac{v(\d\la_1+\d\la_2)}{2\sqrt{2}}\> \big( 
         \pi\!\st\!\pi\!\st\!\s + \s\!\st\!\s\!\st\!\s \big)  
       - \frac{\d\la_1}{4}\> \pi\!\st\! \pi \!\st\!\s\!\st\!\s
       + \frac{\d\la_1 -\d\la_2}{8} \> 
         \pi\!\st\!\s\!\st\!\pi\!\st\!\s }\\[12pt]
\hphantom{S_{\rm ct,br} ~}
  {\ds -\,\frac{\d\la_1 + \d\la_2}{16}\> \big(  
                \pi\!\st\!\pi\!\st\!\pi\!\st\!\pi 
              + \s\!\st\!\s\!\st\!\s\!\st\!\s\big) \bigg\}}
\end{array}
\label{br-ct-action}
\end{equation}
and $\d_1$ and $\d_2$ take the form
\begin{eqnarray}
  & {\ds \d_1 = \d M^2 + \frac{v^2}{2}\> 
           (\d\la_1 + \d\la_2)  } & \label{delta-1} \\
  & {\ds \d_2 = \d M^2 + \frac{3}{2} \>v^2\, 
           (\d\la_1 + \d\la_2)~.  } & \label{delta-2} 
\end{eqnarray}
It is straightforward to check that $S_{\La,\rm br}$ and $S_{\rm
ct,br}$ are both invariant under the $U(1)$ global transformations
(\ref{br-transformation}) and that their Feynman rules are those shown
in figs. 2 and 3.

Introducing real sources $J_\pi$ and $J_\s$ for the fields $\pi$ and
$\s$ through
\begin{displaymath}
  J = \frac{1}{\sqrt{2}}\> \big( J_\pi + i J_\s) \qquad
  J^* = \frac{1}{\sqrt{2}}\> \big( J_\pi - i J_\s)
\end{displaymath}
and substituting in eq. (\ref{generating}), we have for the generating
functional for the Green functions of the fields $\pi$ and $\s$
\begin{equation*}
  Z[J_\pi,J_\s] = e^{G_c[J_\pi,J_\s]} 
    = \int \! [d\pi]\> [d\s] \> \exp\left\{ i S_{\La,\rm br}
    + i S_{\rm ct,br} + i\idx \Big[ J_\pi \pi 
            - J_\s \big( \s + \sqrt{2}\,v\big) \Big] \right\} ~.
\end{equation*}
To obtain the Ward identity that controls the global $U(1)$ symmetry
at the quantum level, we follow the usual method: make the change
(\ref{br-transformation}) in the integral that defines
$Z[J_\pi,J_\s]$, note that $S_{\La,\rm br}$ and $S_{\rm ct, br}$
remain invariant under such a change and define the effective action
$\gm[\pi,\s]$ as the Legendre transform of $W[J_\pi,J_\s]$. This
yields the identity
\begin{equation}
  \idx \left( \s\frac{\d\gm}{\d\pi} - \pi\frac{\d\gm}{\d\s} \right)
     = -\sqrt{2} \idx \frac{\d\gm}{\d\pi} ~ .
\label{br-WI}
\end{equation}
If we denote by $\gm^{(n,m)}(x_1,\ldots,x_n;y_1,\ldots,y_m)$ the 1PI
Green function of $n$ $\pi$-fields and $m$ $\s$-fields, the
effective action can be written as
\begin{displaymath}
\begin{array}{l}
  {\ds \gm[\pi,\s] = \sum_{n,m=1}^\infty \frac{1}{n!\,m!} 
    \int d^4x_1\cdots d^4x_n\, d^4y_1\cdots d^4y_m\> }\\[12pt]
\hphantom{\gm[\pi,\s]\,}
    \times \pi(x_1)\cdots \pi(x_n)\> \s(y_1)\cdots \s(y_m) \>
    \gm^{(n,m)}(x_1,\ldots,x_n;y_1,\ldots,y_m) ~.
\end{array}
\end{displaymath}
Substituting this in eq. (\ref{br-WI}) and going to momentum space, we
obtain the following set of Ward identities for the 1PI Green
functions:
\begin{equation}
\begin{array}{l}
   m\,\gm^{(n+1,m-1)}(p_1,\ldots,p_n,q_m;q_1,\ldots,q_{m-1}) \\
   \qquad -\, n\,\gm^{(n-1,m+1)}(p_1,\ldots,p_{n-1};
                      q_1,\ldots,q_m,q_{m+1}) \\
   \qquad\qquad   = -\sqrt{2}\,v\,
         \gm^{(n+1,m-1)}(p_1,\ldots,p_n,0;q_1,\ldots,q_m)~.
\end{array}
\label{br-1PI-WI}
\end{equation}
It is important to note 
the zero momentum insertion on r.h.s. of the identities, 
since it
will play a key part in our analysis in section 4. The same comments
made for the symmetric phase concerning the quantum theory apply
here. Namely, for the broken phase of the quantum theory to exist, one
must make sure that it is possible to take order by order in
perturbation theory the counterterms so as to render the limit
$\La\to\infty$ of all 1PI Green functions finite, while preserving the
Ward identities. In this section we show that this is possible at one
loop only if $\la_2=0$.

The 1PI Green functions with UV divergences for $\La\to\infty$ in
their planar parts are, in the notation introduced above,
\begin{equation}
\begin{array}{ll}
  \gm^{(0,1)}(0) &  \\
  \gm^{(2,0)}(p) & \\
  \gm^{(0,2)}(q) &  \\
  \gm^{(2,1)}(p_1,p_2;q)    & p_1+p_2+q=0 \\
  \gm^{(0,3)}(q_1,q_2,q_3)  & q_1+q_2+q_3=0 \\
  \gm^{(4,0)}(p_1,p_2,p_3,p_4) \qquad  & p_1+p_2+p_3+p_4=0 \\ 
  \gm^{(2,2)}(p_1,p_2;q_1,q_2)  & p_1+p_2+q_1+q_2=0 \\ 
  \gm^{(0,4)}(q_1,q_2,q_3,q_4)  & q_1+q_2+q_3+q_4=0 ~.
\end{array}
\label{list}
\end{equation}
By the UV/IR mixing argument, these are also the only 1PI the Green
functions whose nonplanar parts may develop singularities at
$\La\to\infty$ when $\theta^{\m\n}\to 0$.  According to
eq. (\ref{br-1PI-WI}), these functions satisfy the Ward identities
\begin{eqnarray}
    \gm^{(0,1)}(0) & \igual & \sqrt{2}\,v\, \gm^{(2,0)}(0) 
              \label{WI-1} \\
    \gm^{(2,0)}(p) - \gm^{(0,2)}(p) & \igual & - \sqrt{2}\,v\,
             \gm^{(2,1)}(p,0;-p)   \label{WI-2} \\
    \,2\gm^{(2,1)}(p,q_1;q_2)  - \gm^{(0,3)}(p,q_1,q_2) & \igual &
         - \sqrt{2}\,v\, \gm^{(2,2)}(p,0;q_1,q_2)  \label{WI-3} \\
    3\,\gm^{(2,1)}(p_1,p_2;p_3) & \igual &  \sqrt{2}\,v\,
         \gm^{(4,0)}(p_1,p_2,p_3,0)   \label{WI-4} \\
    3\,\gm^{(2,2)}(p_1,q_1;q_2,q_3) - \gm^{(0,4)}(p_1,q_1,q_2,q_3) 
         & \igual & - \sqrt{2}\,v\, \gm^{(2,3)}(p_1,0;q_1,q_2,q_3)  
          \label{WI-5}  \\
    \gm^{(4,0)}(p_1,p_2,p_3,q) - 3\, \gm^{(2,2)}(p_1,p_2;p_3,q)
       & \igual & - \sqrt{2}\,v\, \gm^{(4,1)}(p_1,p_2,p_3,0;q) ~. 
         \label{WI-6}
\end{eqnarray}
We first look at $\gm^{(0,1)}(0)$. At one loop, it is given by
\begin{equation}
  \epsfig{file=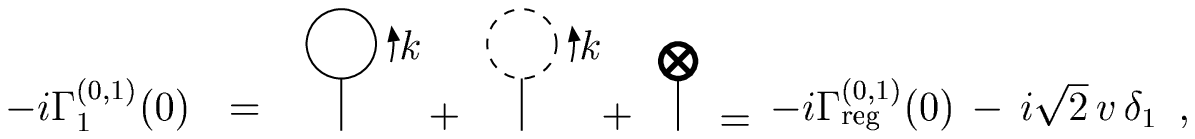}
\label{1-sigma}
\end{equation}
where the regularized contribution $ -i\gm^{(0,1)}_{\rm reg}(0)$ has
the form 
\begin{equation}
    -i\gm^{(0,1)}_{\rm reg}(0) = -i\gm_{\rm P}^{(0,1)}(0) = 
    \frac{v\,(\la_1+\la_2)}{2\sqrt{2}}
      \idk\, \bigg[\, \frac{3}{D_\s(k)} +\frac{1}{D_\pi(k)} \,\bigg] 
\label{1-sigma-integral}
\end{equation}
and we have defined
\begin{eqnarray}
    & {\ds D_\pi(k) = k^2\Big(1-\frac{\ds k^2}{\ds \La^2}\Big)^n } & 
      \label{D-pi} \\
    & {\ds D_\s(k) = k^2\Big(1-\frac{\ds k^2}{\ds \La^2}\Big)^n 
      - 2M^2 ~.} & \label{D-sigma}  
\end{eqnarray}
Note that $i\gm^{(0,1)}_{\rm reg}(0)$ is purely planar and is
quadratically divergent at $\La\to\infty$. To compute its large $\La$
limit we use eqs. (\ref{Ipi-p}) and (\ref{Isigma-p}) in the Appendix 
and obtain
\begin{equation}
   -i\gm^{(0,1)}_{\rm reg}(0) = -i\gm^{(0,1)}_{\rm P}(0) ~~\uvlim~~
        - \frac{i}{16\pi^2} \sqrt{2}v\,(\la_1+\la_2)\>
          \bigg\{ \frac{\La^2}{n-1}
       - \frac{3}{2}\>M^2 \bigg[ \ln\!\Big(\frac{\La^2}{2M^2}\Big) 
       - f_0 \bigg] \bigg\} ~,
\label{1-sigma-planar}
\end{equation}
with $f_0$ as in eq. (\ref{f0}). It follows that, for $\gm^{(0,1)}(0)$
to be finite, $\d_1$ must be modulo finite terms
\begin{equation}
   \d_1  = - \frac{\la_1+\la_2}{16\pi^2}\>\
     \bigg[ \frac{\La^2}{n-1}
    - \frac{3}{2}\>M^2\>\ln\!\Big(\frac{\La^2}{2M^2}\Big) \bigg] ~.
\label{value-delta-1-1}
\end{equation}

Next we consider the pion selfenergy $\gm^{(2,0)}(p)$. At one loop, it
receives contributions from the following 1PI diagrams
\begin{equation}
\epsfig{file=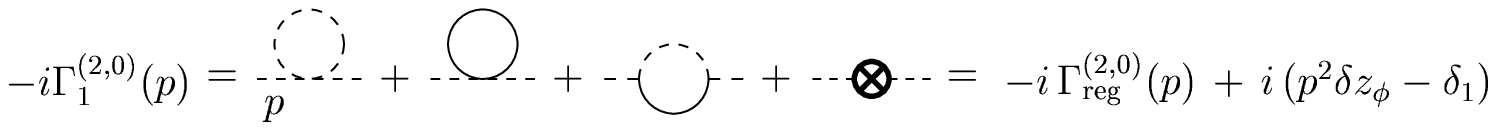}
\label{2-pi}
\end{equation}
Using the Feynman rules, the contribution $-i\gm_{\rm reg}^{(2,0)}(p)$
can be written as the sum
\begin{equation}
  -i\gm_{\rm reg}^{(2,0)}(p) =  -i\gm_{\rm P}^{(2,0)}(p) 
     -i\gm_{\rm NP}^{(2,0)}(p)
\label{2-pi-decom}
\end{equation}
of a planar part $-i\gm_{\rm P}^{(2,0)}(p)$ and a nonplanar part
$-i\gm_{\rm NP}^{(2,0)}(p)$, given by
\begin{equation}
\begin{array}{l}
   {\ds 
    \bigg\{ \begin{array}{c} -i\gm_{\rm P}^{(2,0)}(p) \\
           -i\gm_{\rm NP}^{(2,0)}(p) \end{array} \bigg\} =
    \frac{1}{4} \idk \Bigg[
    \frac{\la_1+\la_2}{D_\pi(k)} ~\PNPo }\\[16pt]
\hphantom{\qquad\qquad}
   {\ds +\, \frac{1}{D_\s(k)} ~ \PNPt 
   + \frac{2M^2(\la_1+\la_2)}
            {D_\pi(k+p)\, D_\s(k)} ~ \PNPtr  \Bigg]} ~ .
\end{array}
\label{20-p-np}
\end{equation}
At $\La\to\infty$, the planar contribution diverges, while the
nonplanar contribution remains finite if $\theta^{\m\n} \neq 0$
\cite{Hayakawa}. From eqs. (\ref{Ipi-p}), (\ref{Isigma-p}) and
(\ref{Isigmapi-p}) we obtain for the large $\La$ limit of the planar
contribution
\begin{equation}
\begin{array}{l}
  {\ds -i\gm_{\rm P}^{(2,0)}(p) \uvlim -\frac{i}{16\pi^2} \>
    \bigg\{ \Big( \la_1+\frac{\la_2}{2}\Big) \, \frac{\La^2}{n-1} 
        }\\[12pt]
\hphantom{ -i\gm_{\rm P}^{(2,0)}(p) \uvlim \>}
  {\ds -\> \frac{3\la_1+\la_2}{2}\>M^2\>\bigg[\,
              \ln\Big(\frac{\La^2}{2M^2}\Big) - f_0 \bigg] 
       + \frac{\la_1+\la_2}{2} \>M^2\,f(p^2)\,\bigg\} ~,}
\end{array}
\label{2-pi-planar}
\end{equation}
where $f(p^2)$ has the form
\begin{equation}
  f(p^2) = 1 - \Big(1 - \frac{2M^2}{p^2}\Big)\,
       \ln\Big( 1 - \frac{p^2}{2M^2}\Big) ~.  
\label{fp}
\end{equation}
Since the UV divergences in the pion selfenergy are those in its
planar part and are given by
eq. (\ref{2-pi-planar}), for the selfenergy to be finite, $\d z_\phi$
and $\d_1$ must be modulo finite terms
\begin{eqnarray}
  & \d z_\phi =  0 &   \\[9pt] \label{value-delta-z}
  & {\ds \d_1 = -\frac{i}{16\pi^2} \>
   \bigg[ \Big( \la_1+\frac{\la_2}{2}\Big) \, \frac{\La^2}{n-1}
     - \frac{3\la_1+\la_2}{2}\> M^2
     \ln\Big(\frac{\La^2}{2M^2}\Big)\bigg] ~.} &
\label{value-delta-1-2}
\end{eqnarray}
Eqs. (\ref{value-delta-1-1}) and (\ref{value-delta-1-2}) imply
\begin{equation}
   \la_2=0 ~.
\label{condition}
\end{equation}
In other words, if $\la_2\neq 0$, there are no counterterms that
consistently subtract the UV divergences in $-i\gm_1^{(0,1)}(0)$ and
$-i\gm_1^{(2,0)}(p)$. Note that the structure of the counterterms in $S_{\rm
ct,br}$, and in particular of those for $-i\gm^{(0,1)}(0)$ and
$-i\gm^{(2,0)}(p)$, results from demanding global $U(1)$ invariance,
so the condition $\la_2=0$ is a requirement of global $U(1)$
invariance.

We now set $\la_2=0$ and compute the UV divergences in the other
Green functions on the list (\ref{list}). Every 1PI Green function
$\gm^{(m,n)}$ on this list is at one loop the sum
\begin{equation}
   \gm^{(m,n)} = \gm_{\rm P}^{(m,n)} + \gm_{\rm NP}^{(m,n)}
        + \gm_{\rm ct}^{(m,n)}~.
\label{decom-1}
\end{equation}
of three terms.  The terms $\gm_{\rm P}^{(m,n)}$ and $\gm_{\rm
NP}^{(m,n)}$ collect the planar and nonplanar contributions of the
corresponding 1PI diagrams formed with the Feynman rules for
$S_{\La,\rm br}$, while the term $\gm_{\rm ct}^{(m,n)}$ is the
counterterm contribution provided by $S_{\rm ct,br}$.  At nonvanishing
external momenta, only the planar part $\gm_{\rm P}^{(m,n)}$ becomes
divergent for $\La\to\infty$. Computing these divergences and summing
to them the counterterm contribution we obtain
\begin{eqnarray}
&{\ds -i\gm^{(0,2)}_1(q) ~\uvlim~ -\frac{i\la_1}{16\pi^2} \> 
     \bigg[ \frac{\La^2}{n-1} - \frac{7}{2}\> M^2 
        \ln\Big(\frac{\La^2}{2M^2}\Big) \bigg] 
       +  i\,(q^2\d z_\phi - \d_2) + {\rm f.\,c.} }&
          \label{02-planar} \\[12pt]  
&{\ds -i\gm^{(2,1)}_1(p_1,p_2;q) ~\uvlim~ \frac{iv}{\sqrt{2}} \>
     \cos\Big( \frac{p_1\wedge p_2}{2}\Big) \bigg[ 
       \frac{\la_1^2}{16\pi^2}\> \ln\Big(\frac{\La^2}{2M^2}\Big) 
       - (\d\la_1+\d\la_2) \bigg]  + {\rm f.\,c.} }&
       \label{21-planar}\\[12pt]
&{\ds -i\gm^{(0,3)}_1(q_1,q_2,q_3)  ~\uvlim~  
    \frac{3iv}{\sqrt{2}}\>\cos\Big(\frac{q_1\wedge q_2}{2}\Big) 
    \bigg[ \frac{\la_1^2}{16\pi^2} \> \ln\Big(\frac{\La^2}{2M^2}\Big) 
    - (\d\la_1+\d\la_2) \bigg] +  {\rm f.\,c.} }& 
    \label{03-planar} 
\end{eqnarray}
\begin{equation}
\begin{array}{l}
  {\ds -i\gm_1^{(4,0)}(p_1,p_2,p_3,p_4) ~\> , ~  
      -i\gm_1^{(0,4)}(p_1,p_2,p_3,p_4)  ~\uvlim~ } \\[9pt] 
\hphantom{-i\gm^{(2,2)}(p_1,p_2,p_3,p_4}
  {\ds ~\uvlim~ \frac{i}{2}\> t_\theta(p_1,p_2,p_3) \bigg[ 
    \frac{\la_1^2}{16\pi^2}\> \ln\Big(\frac{\La^2}{2M^2}\Big) 
    - (\d\la_1 +\d\la_2) \bigg] \,+\, {\rm f.\,c.}}
\end{array}
    \label{40-planar} 
\end{equation}
\begin{equation}
\begin{array}{l}
 {\ds -i\gm_1^{(2,2)}(p_1,p_2;q_1,q_2) ~\uvlim  - \frac{i}{2}\>
       \cos\Big(\frac{p_1\wedge q_2 + p_2\wedge q_1}{2}\Big)
       \bigg[ \frac{\la_1^2}{16\pi^2}\> 
       \ln\Big(\frac{\La^2}{2M^2}\Big) - ( \d\la_1-\d\la_2)  \bigg] }
       \\[12pt]
\hphantom{-i\gm^{(2,2)}(p_1,p_2,p_3,p_4}
 {\ds +\, i \cos\Big(\frac{p_1\wedge p_2}{2}\Big) 
             \cos\Big(\frac{q_1\wedge q_2}{2}\Big)
             \bigg[ \frac{\la_1^2}{16\pi^2}\> 
             \ln\Big(\frac{\La^2}{2M^2}\Big) - \d\la_1 \bigg] 
             + {\rm f.\,c.}~,}
\end{array}
 \label{22-planar}
\end{equation}
where $t_\theta(p_1,p_2,p_3)$ stands for
\begin{eqnarray}
   t_\theta(p_1,p_2,p_3) &\igual & \cosppp \nonumber \\
   & \mas & \cosppm  \nonumber \\
   & \mas & \cospmm ~. \label{factort}
\end{eqnarray}
Note that to calculate the UV divergences of the Green functions
above, among all the 1PI one-loop diagrams that contribute to a given
Green function, we only need to consider those with at most two
internal lines. The reason is that 1PI one-loop diagrams with three or
more internal lines contain at least three propagators and thus their
planar contributions are finite by power counting at $\La\to\infty$.

For $-i\gm_1^{(0,2)}(q)$ in eq. (\ref{02-planar}) to be finite, $\d
z_\phi$ and $\d_2$ must be given, modulo finite terms, by $\d
z_\phi=0$ and
\begin{equation}
   \d_2 = -\frac{\la_1}{16\pi^2} \> 
     \bigg[ \frac{\La^2}{n-1} - \frac{7}{2}\> M^2 
        \ln\Big(\frac{\La^2}{2M^2}\Big) \bigg] ~ .
\label{value-delta-2}
\end{equation}
In turn, modulo finite terms, eqs.  (\ref{delta-1}), (\ref{delta-2}),
(\ref{value-delta-1-1}) and (\ref{value-delta-2}) yield for $\d M^2$
and $\d\la_1+\d\la_2$
\begin{eqnarray}
   & {\ds \d M^2 =  -\frac{\la_1}{16\pi^2} \left[ \frac{\La^2}{n-1} 
      - \frac{M^2}{2} \ln\Big(\frac{\La^2}{2M^2}\Big) \right] } &
    \label{br-delta-m}\\[9pt]
   & {\ds \d\la_1 +\d\la_2 = \frac{\la_1^2}{16\pi^2}\> 
       \ln\Big(\frac{\La^2}{2M^2}\Big) ~.} & \nonumber 
\end{eqnarray}
The latter equation and (\ref{22-planar}) imply
\begin{eqnarray} 
   & {\ds \d\la_2=0} & \label{br-delta-la2} \\[6pt]
    & {\ds \d\la_1 = \frac{\la_1^2}{16\pi^2}\> 
       \ln\Big(\frac{\La^2}{2M^2}\Big) ~.} & \label{br-delta-la1}
\end{eqnarray}
To determine the finite terms in $\d M^2,\,\d z_\phi$ and $\d\la_1$,
three renormalization conditions should be specified.

\section{The broken phase II: Ward identities}

In this section we rederive the condition $\la_2=0$ from the Ward
identities (\ref{WI-1})-(\ref{WI-6}). So let us assume that $\la_2\neq
0$ and recall that the identities hold for all $\La,\,\la_1,\,\la_2$
and all $\d_1,\,\d_2,\,\d\la_1,\,\d\la_2$. With this in mind we look
at the identity (\ref{WI-1}). Using the expressions for
$\gm_1^{(0,1)}(0)$ and $\gm_1^{(2,0)}(p)$ in eqs. (\ref{1-sigma}) and
(\ref{2-pi}), the terms with $\d_1$ cancel and we are left with
\begin{equation}
   \gm_{\rm P}^{(0,1)}(0) = \sqrt{2}\, v \left[ 
      \gm^{(2,0)}_{\rm P}(0) + \gm^{(2,0)}_{\rm NP}(0) \right] ~.
\label{WI-1-pnp}
\end{equation}
The contribution $ \gm_{\rm P}^{(0,1)}(0)$ on the l.h.s. is given in
eq. (\ref{1-sigma-integral}), while for $\gm^{(2,0)}_{\rm P}(0)$ and
$\gm^{(2,0)}_{\rm NP}(0)$ on the r.h.s. we have from eq.
(\ref{20-p-np}) that
\begin{eqnarray}
  & {\ds  -i\gm_{\rm P}^{(2,0)}(0) = \frac{1}{4} \idk \Bigg[
       \frac{\la_1+\la_2}{D_\pi(k)} + \frac{3\la_1+\la_2}{D_\s(k)} 
          \Bigg] } & \label{20-p-0}  \\[6pt]
  & {\ds  -i\gm_{\rm NP}^{(2,0)}(0) = \frac{\la_2}{2} \idk \frac{1}
      {D_\s(k)}~.}  & \label{20-np-0} 
\end{eqnarray}
It is clear from eqs. (\ref{1-sigma-integral}), (\ref{20-p-0}) and
(\ref{20-np-0}) that eq. (\ref{WI-1-pnp}) is satisfied. This is no
surprise since, as stated, the Ward identities hold for arbitrary
$\La,\,\la_1,\,\la_2$. The key point is that the identity
(\ref{WI-1-pnp}) holds because there is a contribution
$\gm^{(2,0)}_{\rm NP}(0)$ to the r.h.s. which is proportional to
$\la_2$, diverges at $\La\to\infty$ and is nonplanar. This indicates a
mismatching in the planar $\La\to\infty$ divergent contributions to
both sides of the identity, or equivalently a mismatching in the UV
divergences\footnote{This mismatching was calculated explicitly in
section 3 [see eqs. (\ref{1-sigma-planar}) and
(\ref{2-pi-planar})]. The argument given here precisely avoids
computing it.}. To subtract the UV divergences, we then have to add
different counterterms to the right and left hand sides, in
contradiction with the statement that the counterterms satisfy the
Ward identities for arbitrary $\la_1$ and $\la_2$.  Hence, to have a
consistent subtraction procedure, we must get rid of the unwanted
divergent contribution $\gm^{(2,0)}_{\rm NP}(0)$, and this implies
taking $\la_2=0$. Note that after setting $\la_2=0$ we are left with
\begin{equation*}
   \lim_{\La\to\infty}\>\gm_{\rm NP}^{(2,0)}(0) = 0~.
\end{equation*}

The argument just given generalizes to the other identities as
follows.  The invariance for arbitrary $\d_1,\,\d_2,\,\d\la_1$ and
$\d\la_2$ of $S_{\rm ct,br}$ under global $U(1)$ transformations
implies that the counterterms in fig. 3 satisfy the Ward
identities. This means that the counterterm contributions to both
sides of the identities cancel, so the identities become relations
among planar and nonplanar parts of Green functions like that in
eq. (\ref{WI-1-pnp}). As $\La\to\infty$, the planar contributions to
the l.h.s. of these relations become singular, while the nonplanar
contributions remain finite. Thus, the divergences that arise for
$\La\to\infty$ on the l.h.s. are of planar type. These divergences
must be matched by only planar divergences on the r.h.s.; otherwise
the UV divergences on the l.h.s. would not be balanced by the UV
divergences on the r.h.s. and their subtraction would require
different counterterms for each side. If all the $\La\to\infty$
divergent contributions to the r.h.s. are to be planar, the nonplanar
contributions to this side should remain finite for $\La\to\infty$.
This, however, is not granted, since on the r.h.s. of the identities
one of the external momenta, say $p_e$, vanishes 
and the nonplanar contributions given by nonplanar Feynman integrals
with nonplanarity factor $e^{ik\wedge p_e}$ become divergent at
$\La\to\infty$ if $p_e=0$.  Hence, we must find conditions that rid
the r.h.s. of the Ward identities of nonplanar contributions which
for $p_e=0$ become divergent at $\La\to\infty$
%
%
Note that what we have precisely proved in
our analysis above of the identity (\ref{WI-1}) is that the condition
$\la_2=0$ ensures the finiteness of $\gm^{(2,0)}_{\rm NP}(0)$ at
$\La\to\infty$. Setting $\la_2=0$, we have checked that all the
nonplanar contributions to the r.h.s. of the identities
(\ref{WI-2})-(\ref{WI-6}) are finite for arbitrary $\la_1$, so no
further condition is required.

The quantum theory being defined as the large $\La$ limit of the
theory at finite $\La$ implies that the Green function
$\gm_1^{(0,1)}(0)$ on the l.h.s. of the Ward identity (\ref{WI-1})
must be computed at $\La\to\infty$ and the function $\gm_1^{(2,0)}(p)$
on the r.h.s. at $\La\to\infty,\,p\to 0$. In our analysis above of the
identity (\ref{WI-1}), for the r.h.s. we have first set $p=0$ in
$\gm_1^{(2,0)}(p)$ and then sent $\La\to\infty$. Setting $p=0$ led to
eqs. (\ref{20-p-0}) and (\ref{20-np-0}), and sending $\La\to\infty$ to
the discussion that follows them. There is, however, one other way to
compute the renormalized Green function $\gm_1^{(2,0)}(p)$ at
$\La\to\infty,\,p\to 0$; namely, to take $\La\to\infty$ at
nonvanishing $p$ and then send $p$ to zero.  For the quantum theory to
be well defined, both procedures must yield the same result. Let us
see that this is the case. To this end we consider again the Ward
identity (\ref{WI-1} and take first $\La\to\infty$ and then $p\to 0$.
The only contribution to the l.h.s. of the identity is the
$p$-independent planar piece $\gm_1^{(0,1)}(0)$, whose large $\La$
limit gives a divergent contribution (which will be canceled by a
suitable counterterm). The r.h.s., in turn, receives contributions
from $\gm_{\rm P}^{(2,0)}(p)$ and $\gm_{\rm NP}^{(2,0)}(p)$. Taking
$\La\to\infty$ at nonvanishing $p$ in the planar contribution
$\gm_{\rm P}^{(2,0)}(p)$ gives eq. (\ref{2-pi-planar}), and sending in
it $p$ to zero yields a $\La\!$-divergent $p\!$-independent
contribution. Proceeding similarly with the nonplanar contribution
$\gm^{(2,0)}_{\rm NP}(p)$, and after using the results in the
Appendix, we obtain
\begin{equation}
  \lim_{p\to 0}\>\lim_{\La\to\infty} \gm_{\rm NP}^{(2,0)}(p) = 
     - \frac{i\la_2}{16\pi^2}\> \left\{ \frac{2}{p\circ p}
           + M^2\left[ \ln(2M^2 p\circ p) 
             - \ln2 + \ga -\frac{1}{2} \,\right] \right\} ~.
\label{20-np-Lap}
\end{equation}
Summing the planar and nonplanar contributions to the r.h.s. of the
identity, we get a $\La\!$-divergent $p\!$-independent term (which
will eventually be canceled by the appropriate counterterm) and a
singular $p\!$-dependent piece $1/{p\circ p}$ which will not be
cancelled by a counterterm and is not on the l.h.s., since the l.h.s
does not depend on $p$.  To avoid this mismatching of singular
$p$-dependent contributions so that the Ward identity holds, we must
eliminate such $p\!$-dependence from the r.h.s., hence we must take
$\la_2=0$. Furthermore, only after setting $\la_2=0$, the planar
contributions to both sides of the identity, given by
eqs. (\ref{2-pi-planar}) and (\ref{1-sigma-planar}), match and the
counterterm is the same for both sides of the identity (see section
3). Thus, sending $\La\to\infty$ in $\gm^{(2,0)}_{\rm NP}(p)$ and then
$p\to 0$ leads to $\la_2=0$ and gives
\begin{equation*}
   \lim_{p\to 0}\>\lim_{\La\to\infty} \gm_{\rm NP}^{(2,0)}(p) = 0~.
\end{equation*}
We have repeated this analysis for the other Ward
identities (\ref{WI-2})-(\ref{WI-6}) and checked that, for $\la_2=0$
and arbitrary $\la_1$, the Green functions on the r.h.s. are free of
divergences in $p_e\!$ and yield the same nonplanar contributions as
if one first sets $p_e=0$ and then sends $\La\to\infty$, $p_{\rm e}$
denoting the vanishing external momentum.

The difference with the Ward identities for the symmetric phase is the
zero momentum insertion on the r.h.s. of the identities. At
$\La\to\infty$, the zero momentum insertion produces UV divergences
proportional to $\la_2$ that, being nonplanar, can not be locally
subtracted. The condition $\la_2=0$ sets such divergences to zero.
Noting that
\begin{itemize}
\item[~~(1)] the Ward identities hold for all $\La,\,\la_1$ and that 
$\la_2$, and 
\item[~~(2)] the only breakings at $\La\to\infty$ may arise from
divergent contributions, and these preserve the identities if $\la_2=0$,
\end{itemize}
we conclude that the Ward identities hold for $\La\to\infty$ if
$\la_2=0$.  This ensures the one-loop existence of the quantum broken
phase for $\la_2=0$.

\section{The Broken phase III: noncommutative IR divergences and the
Goldstone theorem}

Here we give explicit expressions for the behaviour of the nonplanar
parts of the Green functions in eq. (\ref{list}) at small
$\theta^{\m\n}$ and show that there is no UV/IR duality in the strong
sense. From section 3 we know that $\gm_1^{(0,1)}(0)$ does not have
nonplanar contributions,
\begin{equation*}
   \gm_{\rm NP}^{(0,1)}(0) = 0.
\end{equation*}
The nonplanar part of $\gm_1^{(2,0)}(p)$ is given by
eq. (\ref{20-p-np}) with $\la_2=0$. Using formuli (\ref{Ipi-ir}),
(\ref{Isigma-ir}) and (\ref{Isigmapi-ir}) to calculate its behaviour
for large $\La$ and small $\theta^{\m\n}$, we obtain
\begin{equation}
  \lim_{\theta^{\m\n}\to 0}\>\lim_{\La\to\infty}\,
      \left[ -i\gm_{\rm NP}^{(2,0)}(p)\right] 
        = \frac{i\la_1}{16\pi^2}\> \frac{M^2}{2}\> f(p^2)~, 
\label{20-ir}
\end{equation}
with $f(p^2)$ as in eq. (\ref{fp}). For the nonplanar parts of the
other Green functions on the list (\ref{list}), after some
calculations and using the results in the Appendix, we have 
\begin{eqnarray}
   &{\ds -i\gm^{(0,2)}_{\rm NP}(q) \approx -\frac{i\la_1}{16\pi^2} 
     \>6 M^2 \ln(\theta M^2) } & 
     \label{02-ir} \\
   &{\ds  -i\gm^{(2,1)}_{\rm NP}(p_1,p_2;q) \approx
     -\frac{iv}{\sqrt{2}} \> \cos\Big( \frac{p_1\wedge p_2}{2}\Big)
     \,\frac{\la_1^2}{16\pi^2}\> 3 \ln(\theta M^2) } & 
     \label{21-ir}\\
   &{\ds -i\gm^{(0,3)}_{\rm NP}(q_1,q_2,q_3)  \approx  -\,
     \frac{3iv}{\sqrt{2}}\>\cos\Big(\frac{q_1\wedge q_2}{2}\Big)\,
     \frac{\la_1^2}{16\pi^2}\> 3\,\ln\,(\theta M^2) } &
     \label{03-ir} \\
   &{\ds -i\gm_{\rm NP}^{(4,0)}(p_1,p_2,p_3,p_4) \approx 
       -i\gm_{\rm NP}^{(0,4)}(p_1,p_2,p_3,p_4) \approx -\, \frac{i}{2} 
       \> t_\theta(p_1,p_2,p_3) \,\frac{\la_1^2}{16\pi^2}\,3 
       \ln\,(\theta M^2) } &
       \label{40-ir} \\
   &{\ds -i\gm_{\rm NP}^{(2,2)}(p_1,p_2;p_3,p_4) \approx -\frac{i}{2}\,
      \cos\Big(\frac{p_1\wedge p_4 + p_2\wedge p_3}{2}\Big)\,
      \frac{\la_1^2}{16\pi^2}\> 3 \ln\,(\theta M^2) ~,} &
\label{22-ir}
\end{eqnarray}
where $t_\theta(p_1,p_2,p_3)$ is as in eq. (\ref{factort}).

Comparing these expressions with eqs. (\ref{1-sigma-planar}),
(\ref{2-pi-planar}) and (\ref{02-planar})-(\ref{22-planar}), we see
that the noncommutative IR singularities and the UV divergences can
not be obtained from each other by replacing $\theta^2M^2
\leftrightarrow 1/\La^2$, thus showing that there is no UV/IR duality
in the strong sense. We also note that, unlike the symmetric phase,
there are no quadratic noncommutative IR divergences. Indeed, the
selfenergy of the $\s$-field only contains logarithmic noncommutative
IR divergences, and the selfenergy of the $\pi$-field does not develop
any noncommutative IR singularity at all. It is also clear from the
equations above that the noncommutative IR singularities satisfy the
Ward identities. This is no surprise, since we know from section 4
that the Ward identities hold and taking $\theta^{\m\n}\to 0$ amounts
to setting $\tp_i\to 0$ as external momentum configuration.

We finally want to study if the Goldstone theorem holds at one
loop. To do this, we need the renormalized pion selfenergy. As is
usual in the commutative case, we take as one of the renormalization
conditions that the vacuum expectation value of the field $\s$ remains
equal to its classical value, {\it i.e.}  $\langle \s \rangle =
v$. This is equivalent to $-i\gm_1^{(1,0)}=0$, which together with
eqs. (\ref{1-sigma}) and (\ref{1-sigma-planar}) completely specifies
$\d_1$ as
\begin{equation*}
   \d_1 = - \frac{\la_1}{16\pi^2} \> \bigg\{ \frac{\La^2}{n-1}
       - \frac{3}{2}\>M^2 \bigg[ \ln\!\Big(\frac{\La^2}{2M^2}\Big) 
       - f_0 \bigg] \bigg\} ~.   
\end{equation*}
Substituting this in eq. (\ref{2-pi}), using
(\ref{2-pi-decom})-(\ref{2-pi-planar}) and summing the tree-level and
one-loop contributions, we obtain for the renormalized pion
selfenergy
\begin{equation}
    \gm^{(2,0)}_{\rm R} (p) = p^2 - \frac{\la_1}{16\pi^2} \> 
      \frac{M^2}{2} \>f(p^2) - \gm_{\rm NP}(p)~, 
\label{renormalized}
\end{equation}
where $f(p^2)$ is as in eq. (\ref{fp}) and
\begin{equation*}
   \gm_{\rm NP}(p) = \lim_{\La\to\infty} \> 
                        \gm^{(2,0)}_{\rm NP} (p)
\end{equation*}
is the large $\La$ limit of the nonplanar contribution
$\gm^{(2,0)}_{\rm NP} (p)$ to the pion selfenergy.  To compute
$\gm_{\rm NP}(p)$, we use eqs.  (\ref{Ipi-ir}), (\ref{1-line})
and (\ref{2-line}) in the Appendix for the three terms in
(\ref{20-p-np}) and obtain
%
%
\begin{eqnarray}
   \gm_{\rm NP}(p) &\igual& \frac{\la_1}{16\pi^2} \> 
       \bigg [ \>\frac{1}{p\circ p} -  2M^2\>
       \frac{K_1\Big(\sqrt{2p\circ p\,M^2}\Big)}{\sqrt{2p\circ p\,M^2}}
       \> \bigg] \nonumber \\[6pt]
   &\menos&  \frac{\la_1}{16\pi^2} \frac{M^2}{2} \int_0^\infty 
        \frac{dt}{t} \int_0^1 d\a~
      {\rm exp}\bigg\{ t\,\a(1-\a)\,p^2 - \,2t\a M^2 
           - \frac{p\circ p}{4t}\,\bigg\}~.  \label{dispersion}
\end{eqnarray}
If we define the mass squared as the value of $p^2$ for which the
selfenergy vanishes, to find the pion mass, we must solve the equation
$\gm^{(2,0)}_{\rm R}(p)=0$. Note in this regard that the renormalized pion
selfenergy is a regular function of $p^2,\,M^2$ and $p\circ p$, so the
equation $\gm^{(2,0)}_{\rm R}(p)=0$ may in principle have
$\theta^{\m\n}\!\!$-dependent solutions with $\theta^{\m\n}\!\neq 0$
and $p^\m\neq 0$. To solve $\gm^{(2,0)}_{\rm R}(p)=0$, we proceed by
iteration and, since at tree level the solution is $p^2=0$, we write
$p^2= \la_1 \d p^2 + O(\la_1^2)$. Substituting this in
eq. (\ref{renormalized}) and noting that $\,f(p^2)\to 0\,$ for $\,p^2\to
0$, we are left with
\begin{equation}
    \d p^2 = \gm_{\rm NP}(p^2=0)~.
\label{Goldstone}
\end{equation}
Setting $p^2=0$ in the second line in eq. (\ref{dispersion}) and
performing the integral, it is straightforward to see that the two
lines in eq. (\ref{dispersion}) cancel each other, so that
$\gm_{\rm NP}(p^2=0)=0$. The solution to $\,\gm^{(2,0)}_{\rm R}(p)=0$,
up to order $\la_1$, is then $\,p^2=0$ and the Goldstone theorem is
preserved by one-loop radiative corrections. Note that the fact that
the renormalized pion selfenergy is free of noncommutative IR
singularities is essential for the Goldstone theorem to hold at one
loop. Had the selfenergy developed noncommutative logarithmic IR
singularities, these would have entered the mass as $\ln(\theta M^2)$,
making it ill defined for small $\theta$.

\section{Conclusion and discussion}

We have studied the one-loop renormalizability, the noncommutative IR
singularities and the UV/IR mixing in both the symmetric and the
broken phases of noncommutative global $U(1)$ scalar field theory. We
have considered the general case of two interaction terms in the
classical action, $\la_1\,\phi^*\!\st\phi\st\phi^*\!\st\phi$ and
$\la_2\,\phi^*\!\st\phi^*\!\st\phi\st\phi$, and used as regulator an
invariant cutoff $\La$. For the symmetric phase, we have shown that
the quantum theory exists at one loop for all values of the coupling
constants $\la_1$ and $\la_2$ compatible with perturbation theory, and
that there is no need to take $\la_2=0$.  We have also given explicit
expressions for the noncommutative IR singularities and checked that
UV/IR duality does not hold in its strong form.

As concerns the broken phase, we have seen that the Ward identities
imply that the quantum theory exists at one loop only if $\la_2$
vanishes. This is so because the Ward identities have a zero-momentum
insertion term that for large $\La$ yields UV divergent contributions
proportional to $\la_2$ that can not be locally subtracted. To have a
renormalizable theory, one must get rid of such contributions, and
this requires $\la_2=0$. We have also given explicit expressions for
the noncommutative IR singularities in the 1PI Green functions of the
the broken phase and shown that there is no strong UV/IR duality. The
situation as concerns noncommutative IR singularities, UV/IR duality
and the Ward identities is different to those cases previously studied
in the literature. Consider for example $U(1)$ gauge theory: since
UV/IR duality holds and the UV divergences are consistent with the
Ward identities, the logarithmic noncommutative IR singularities
satisfy the Ward identities. For the case at hand, however, the UV
divergences satisfy the Ward identities, there is no UV/IR duality
and, yet, the noncommutative IR singularities satisfy the Ward
identities.

Comparing the symmetric and the broken phases, we have seen that after
spontaneous symmetry breaking the theory does not have quadratic
noncommutative IR divergences. Furthermore, the pion selfenergy is
free of noncommutative IR singularities of any type, which makes
possible to formulate the Goldstone theorem for all $\theta^{\m\n}$.
Had UV/IR hold, the pion selfenergy would have contained
noncommutative logarithmic IR singularities $\ln(\theta M^2)$ and
these would have spoiled the theorem.  Since the interaction term
$\phi^*\!\st\phi\st\phi^*\!\st\phi$ for which the broken phase makes
sense at one loop is also invariant under local $U(1)$ gauge
transformations, it would be interesting to investigate the
implications of noncommutative IR singularities and UV/IR mixing for
the Goldstone theorem in local models \cite{Petriello} \cite{Liao}.

\section*{Acknowledgment}

The author is grateful to C.P. Mart\'{\i}n for many illuminating
conversations and for reading the manuscript. He also acknowledges
financial support from CICyT, Spain through grant No. PB98-0842.

\section*{Appendix}
{\renewcommand{\theequation}{A.\arabic{equation}}\setcounter{equation}{0}

In the computations we have performed in sections 2 to 5 we have
encountered the following integrals:
\begin{eqnarray}
  I_\pi(q) &\igual & \idk \frac{e^{i q\wedge k}}{D_\pi(k)} 
        \nonumber \\[6pt]
  I_\s(q,M) &\igual & \idk \frac{e^{i q\wedge k}}{D_\s(k)} 
         \nonumber \\[6pt]
  I_{\pi\pi}(q,p) &\igual & \idk 
    \frac{e^{i q\wedge k}}{D_\pi(k)\>D_\pi(k+p)} 
         \label{ap-list} \\[6pt]
  I_{\s\pi}(q,p,M) &\igual & \idk 
    \frac{e^{i q\wedge k}}{D_\s(k)\>D_\pi(k+p)} \nonumber\\[6pt] 
  I_{\s\s}(q,p,M) &\igual& \idk 
        \frac{e^{i q\wedge k}}{D_\s(k)\>D_\s(k+p)} ~.
        \nonumber
\end{eqnarray}
We are interested in their large $\La$ limit. To
compute it, we proceed as follows. We first Wick rotate to euclidean
space, make the change $k\to k\La$ and define $\hp^\m\equiv
{p^\m}/{\La}$ and $\hm\equiv {M}/{\La}$.  The integrals above then
become functions of the dimensionless variables $\tq^\m\La,\,\hp^\m$
and $\hm$. Next we use algebraic identities like 
\begin{equation*} 
  \frac{1}{1+(k+\hp)^2} = \frac{1}{1+k^2} \left[ 1 -
    \frac{\hp^2 + 2\hp k}{1+(k+\hp)^2} \right]
\end{equation*}
or 
\begin{equation*}
  \frac{1}{k^2 {(1+k^2)}^n + 2\hm^2} = \frac{1}{k^2+2\hm^2}
    \left[\, 1 - \sum_{r=1}^n \,\noverr\>
        \frac{(k^2)^{r+1}}{k^2 {(1+k^2)}^n + 2\hm^2}\, \right]
\end{equation*}
to decompose every integral in a sum of integrals, whose limit
$\La\to\infty$ we study employing Lebesgue's dominated convergence
theorem.  Finally we use Schwinger parameters to compute the integrals
that give nonvanishing contributions at $\La\to\infty$ and rotate back
to Minkowski spacetime. Following this procedure we obtain for $q=0$
\begin{equation}
   I_\pi(0) ~\uvlim~ - \frac{i}{16\pi^2}\>\frac{\La^2}{n-1}   
\label{Ipi-p}
\end{equation}
\begin{eqnarray}
   &{\ds I_\s(0,M) ~\uvlim~ - \frac{i}{16\pi^2}\,\left\{ 
     \frac{\La^2}{n-1} - 2M^2\,\left[\, 
     \ln\Big(\frac{\La^2}{2M^2}\Big) - f_0 \right] \right\} }&
     \label{Isigma-p} \\
   &{\ds I_{\pi\pi}(0,p) ~\uvlim~ \frac{i}{16\pi^2}\, 
      \bigg[ - \ln\!\Big(\!\!-\!\frac{p^2}{\La^2}\Big) + 1    
      - \sum_{r=1}^{2n-1} \, \frac{1}{\gm(r)} \bigg] } &
   \nonumber \\
   &{\ds I_{\s\pi}(0,p,M) ~\uvlim~ \frac{i}{16\pi^2}\, \bigg[ 
     \ln\!\Big(\frac{\La^2}{2M^2}\Big) + f(p^2) - f_0 \bigg] }&
     \label{Isigmapi-p} \\
   &{\ds I_{\s\s}(0,p,M) ~\uvlim~ \frac{i}{16\pi^2}\,\bigg[
   \ln\!\Big(\frac{\La^2}{2M^2}\Big) - g(p^2) - f_0 \bigg] ~,} 
  \nonumber 
\end{eqnarray}
where $f_0$ and $f(p^2)$ are as in eqs. (\ref{f0}) and (\ref{fp}) and
$g(p^2)$ reads
\begin{equation*}
  g(p^2) = 2 - \sqrt{1-8M^2/p^2}\> \ln\!\left( 
          \frac{\sqrt{1-8M^2/p^2} +1}{\sqrt{1-8M^2/p^2} -1} \right) ~.
\end{equation*}
For $q\neq 0$, the results for $I_\pi$ and $I_\s$ are relatively
simple,
\begin{eqnarray}
   &{\ds I_\pi(q) ~\uvlim~  - \frac{i}{4\pi^2}\>\frac{1}{q\circ q}  }&
     \label{Ipi-ir}\\
   &{\ds I_\s(q,M) ~\uvlim~ \frac{iM^2}{2\pi^2}\> 
     \frac{K_1(\sqrt{2\,q\circ q M^2})}{\sqrt{2\,q\circ q M^2}}~, }&
     \label{1-line}
\end{eqnarray}
whereas for $I_{\pi\pi},\>I_{\s\pi}$ and $I_{\s\s}$ we
have
%
%
\begin{equation}
   \left. \begin{array}{r} I_{\pi\pi}(q,p)\\
                            I_{\s\pi}(q,p,M)\\
                            I_{\s\s}(q,p,M) \end{array}\right\}
   ~\uvlim~ \frac{i}{16\pi^2} \int_0^\infty 
   \frac{dt}{t} \int_0^1 \!d\a~ {\rm exp}\bigg[ t\,\a(1-\a)\,p^2 
      - \,2t \epsilon M^2 - \frac{q\circ q}{4t} - i\a q\wedge p\,\bigg] ~,
\label{2-line}
\end{equation}
with
\begin{equation}
   \epsilon =  \left\{ \begin{array}{ll} 
                       0 &\quad {\rm for}~I_{\pi\pi} \\
                       \a &\quad {\rm for}~I_{\s\pi} \\
                       1  &\quad {\rm for}~I_{\s\s} ~.
                       \end{array} \right.
\label{param}                    
\end{equation}
To study the noncommutative IR singularities in the 1PI Green
functions we need only the expressions at
$\La\to\infty,\,\theta^{\m\n}\to 0$. They can be easily computed and
turn out to be
\begin{eqnarray}
   &{\ds \lim_{q\to 0}\> \lim_{\La\to\infty} ~I_\s(q,M) =
     \frac{iM^2}{a\pi^2}\,\left[ \frac{1}{q\circ q\,M^2} 
       -\frac{1}{2}\,\ln(2q\circ q\,M^2) + \ga - \ln 2 - \frac{1}{2}\,
          \right]} 
   \label{Isigma-ir}\\
   &{\ds \lim_{q\to 0}\> \lim_{\La\to\infty} ~I_{\pi\pi}(q,p) =
     - \frac{i}{16\pi^2}\, \left[ \ln(-\!q\circ q\, p^2)
        + 2\ln2 -2\ga +1 \right] }&
   \label{Ipipi-ir}\\
   &{\ds   \lim_{q\to 0} \lim_{\La\to\infty}~I_{\s\pi}(q,p,M) =
      - \frac{i}{16\pi^2}\,\bigg[ \, \ln(2\,q\circ q M^2) - f(p^2)
     - 2,(\ln2 -\ga) - 1  \bigg] }&
    \label{Isigmapi-ir}\\
   &{\ds   \lim_{q\to 0}\> \lim_{\La\to\infty} ~I_{\s\s}(q,p,M) =
      - \frac{i}{16\pi^2}\,\bigg[ \, \ln(2\,q\circ q M^2) - g(p)
      -\ln 2 + \ga -1 \bigg] ~.} & 
   \nonumber 
\end{eqnarray}
Note {\it e.g.} that substitution of eqs. (\ref{Ipi-ir}),
(\ref{Isigma-ir}) and (\ref{Isigmapi-ir}) in (\ref{20-p-np}) yields
eq. (\ref{20-np-Lap}).

\newpage

\vspace{20pt}
\begin{center}
   \epsfig{file=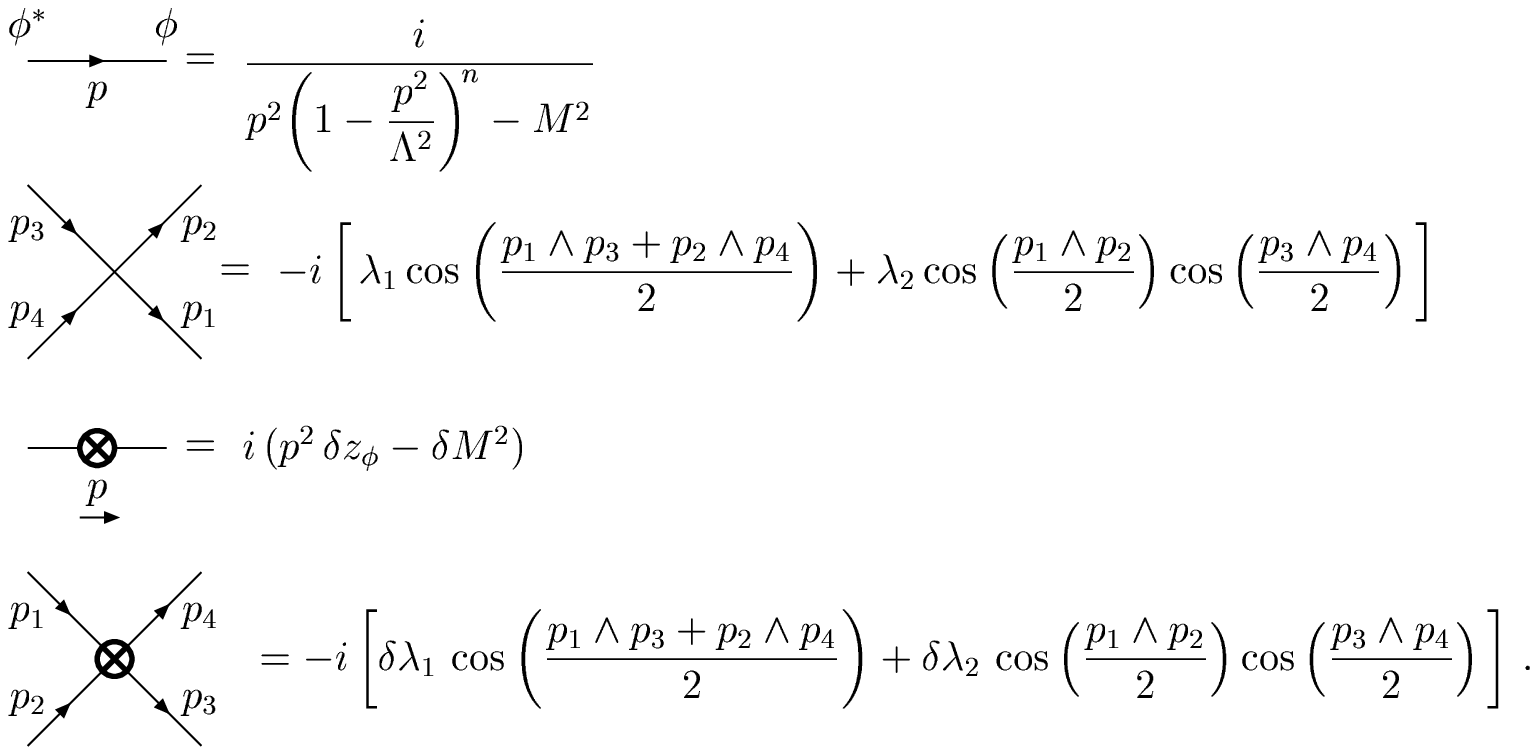}\\[25pt]
{\sl Figure 1: Feynman rules for $S_{\La,\rm sym}$ and 
    $S_{\rm ct,sym}$.}
\end{center}
\newpage
\begin{center}
   \epsfig{file=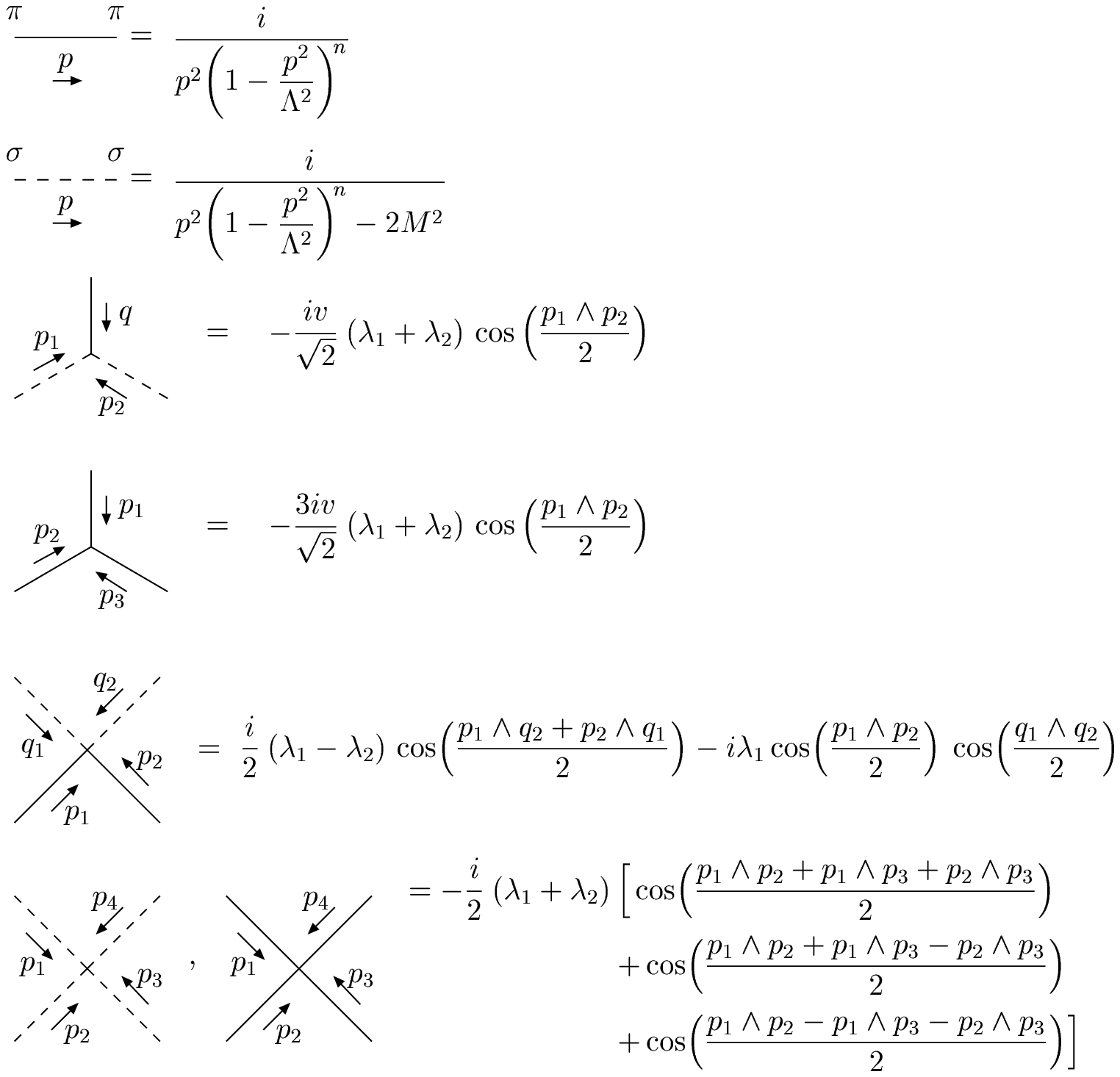}\\[10pt]
{\sl Figure 2: Feynman rules for $S_{\La,\rm br}$.}
\end{center}
\newpage
\begin{center}
   \epsfig{file=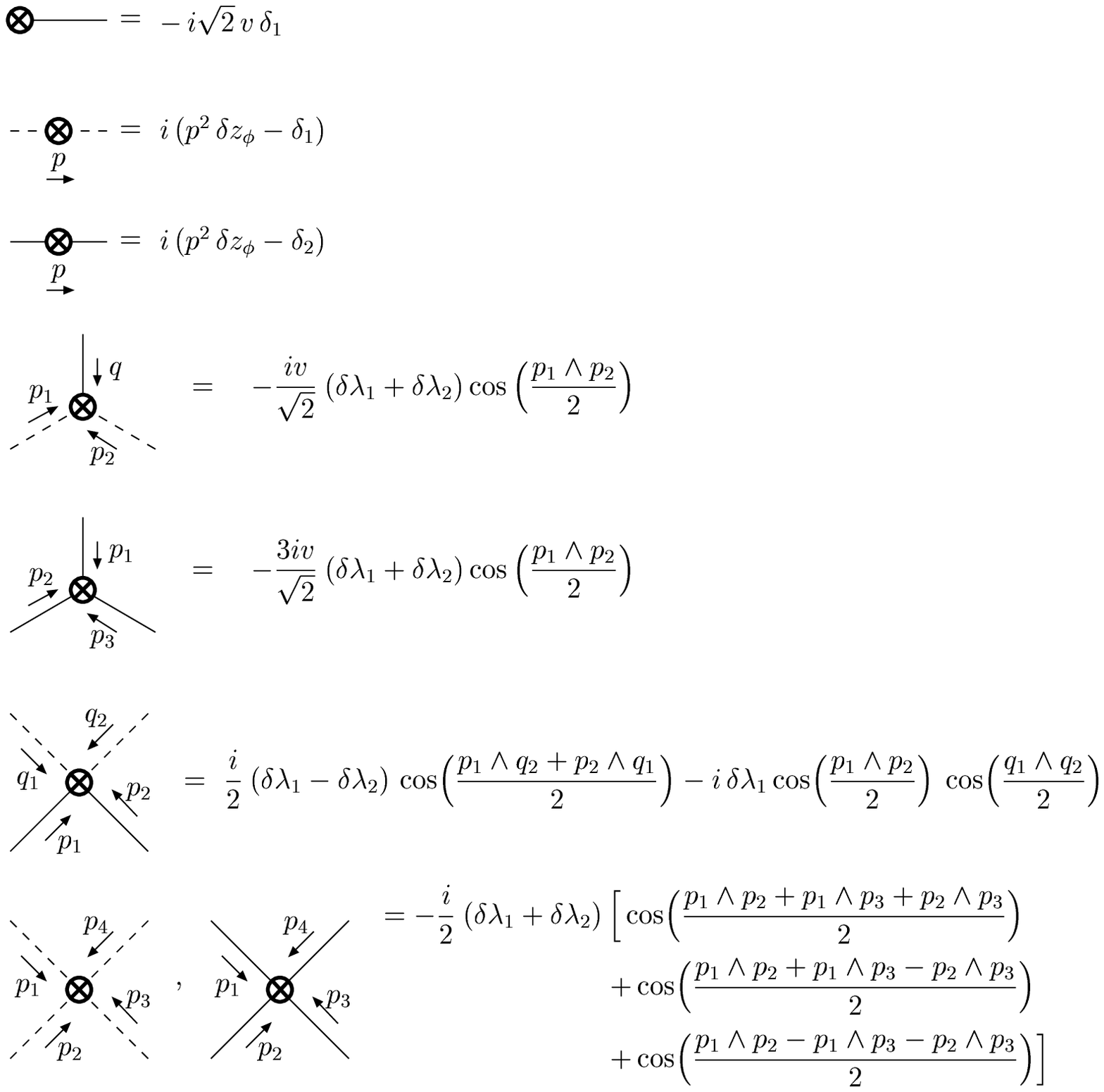}\\[10pt]
{\sl Figure 3: Feynman rules for $S_{\rm ct, br}$.}
\end{center}


\begin{thebibliography}{99}
%
\bibitem{reviews} 
For recent reviews on noncommutative field theory, see\\ 
%
  M.R. Douglas and N.A. Nekrasov, {\it Noncommutative field theory},
  Rev. Mod. Phys. {\bf 73} (2002) 977.\\
%
  R.J. Szabo, {\it Quantum field theory on noncommutative spaces},
  IEEE Trans. Nucl. Sci. 48 (2001) pp. mult. p.\\
%
I. Ya. Aref'eva, D.M. Belov, A.A. Giryavets, A.S. Koshelev and
P.B. Medvedev, {\it Noncommutative field theories and (super)string
field theories}, {\tt hep-th/0111208}.
%
\bibitem{Minwalla} 
  S. Minwalla, M. V. Raamsdonk and N. Seiberg, {\it
  Noncommutative perturbative dynamics}, JHEP {\bf 0002} (2000) 020.
%
\bibitem{Hayakawa}
   M. Hayakawa, {\it Perturbative analysis of IR aspects of
   noncommutative QED on ${\mathbf R}^4$}, Phys. Lett. {\bf B478}
   (2000) 394.
%
\bibitem{Matusis} 
   A. Matusis, L. Susskind and N. Toumbas, {\it The IR/UV connection
   in the noncommutative gauge theories}, JHEP {\bf 0012} (2000)
   002. 
%
\bibitem{Martin}
    C.P. Mart\'{\i}n and F. Ruiz Ruiz, {\it Paramagnetic dominance,
    the sign of the beta function and UV/IR mixing in noncommutative
    $U(1)$}, Nucl. Phys. {\bf B597} (2001) 197.
%
\bibitem{vac}
   F. Ruiz Ruiz, {\it Gauge-fixing independence of IR divergences in
   noncommutative $U(1)$, perturbative tachyonic instabilities and
   supersymmetry}, Phys. Lett. {\bf B502} (2001) 274.
%
\bibitem{Andreev}
   O. Andreev and H. Dorn, {\it Diagrams of noncommutative $\phi^3$ 
   theory from string theory}, Nucl. Phys. {\bf B583} (2000) 145.\\
%
   A. Bilal, C-S. Chu and R. Russo, {\it String theory and noncommutative
   field theories at one loop}, Nucl. Phys. {\bf BB582} (2000) 65.\\
%
   Y. Kiem and S. Lee, {\it UV/IR mixing in noncommutative field
   theory via open string loops}, Nucl. Phys. {\bf B586} (2000) 303.\\
%
   H. Liu and J. Michelson, {\it Stretched strings in noncommutative
   field theory}, Phys. Rev. {\bf D62} (2000) 066003.\\
%
   J. Gomis, M. Kleban, T. Mehen, M. Rangamani and S. Shenker, {\it
   Non-commutative gauge dynamics from the string worldsheet}, JHEP
   {\bf 0008} (2000) 011.\\
%
   A. Armoni and E. L\'opez, {\it UV/IR mixing via closed strings and
   tachyonic instabilities}, {\tt hep-th/0110113}.
%
\bibitem{2loop} 
   An explicit proof of two-loop renormalizability for $\la\phi^4$ is
   given in I. Ya. Aref'eva, D.M. Belov and A.S. Koshelev, {\it
   Two-loop diagrams in noncommutative $\varphi^4_4$ theory},
   Phys. Lett. {\bf B476} (2000) 431.\\
%
   Paticular types of $n$-loop diagrams in $\la\phi^4$ have been
   studied in I. Chepelev and R. Roiban, {\it Convergence theorem for
   noncommutative Feynman graphs and renormalization}, JHEP {\bf 0103}
   (2001) 001.\\
%
   A wilsonian approach can be found in L. Griguolo and M. 
   Pietroni  , {\it Hard noncommutative loops resumation}, 
   {\tt hep-th/0102070}.
%
\bibitem{Girotti}
   H.O. Girotti, M. Gomes, V. O Rivelles and A.J. da Silva, {\it A
   consistent noncommutative field theory: the Wess-Zumino model},
   Nucl .Phys. {\bf B587 } (2000) 299.\\
   H.O. Girotti, M. Gomes and Yu. Petrov, {\it The three-dimensional 
   noncommutative nonlinear sigma model in superspace},
   Phys. Lett. {\bf B521} (2001) 119.
%
\bibitem{Mehen} 
   J. Gomis, T. Mehen and M.B. Wise, {\it Quantum Field theories with
   compact noncommutative extra dimensions}, JHEP {\bf 0008} (2000)
   029.\\ 
%
   J. Gomis, K. Landsteiner and E. L\'opez, {\it Non-relativistic
   noncommutative field theory and UV/IR mixing}, Phys.Rev. {\bf D62}
   (2000) 105006.\\
%
   K. Landsteiner, E. L\'opez and M.H.G. Tytgat, {\it Instability of
   noncommutative SYM theories at finite temperature}, JHEP {\bf 0009}
   (2000) 027.
%
\bibitem{gauge}
   C.P. Mart\'{\i}n and D. S\'anchez-Ruiz, {\it The on-loop UV
   structure of $U(1)$ Yang-Mills theory in noncommutative ${\mathbf
   R}^4$}, Phys. Rev. Lett. {\bf 83} (1999) 476.\\
%
   M. Sheikh-Jabbari, {\it Renormalizability of the supersymmetric
   Yang-Mills theories on the noncommutative torus}, JHEP {\bf 9906}
   (1999) 015.\\
%
   T. Krajewski and R. Wulkenhaar, {\it Perturbative quantum gauge
   fields on the noncommutative torus}, J. Mod. Phys. {\bf A15} (2000)
   1011.\\
%
   A. Armoni,  {\it Comments on perturbative dynamics of noncommutative 
   Yang-Mills theory}, Nucl. Phys. {\bf 593} (2001) 229.\\
%
   C.P. Mart\'{\i}n and D. S\'anchez-Ruiz, {\it The BRS invariance of
   noncommutative $U(N)$ Yang-Mills theory at the one loop level},
   Nucl. Phys.  {\bf B598} (2001) 348.
%
\bibitem{Gomis}
   J. Gomis and T. Mehen, {\it Space-time noncommutative field theories
   and unitarity}, Nucl Phys. {\bf B591} (2000) 265.\\
%
   L. Alvarez-Gaum\'e and J.L.F. Barb\'on , {\it Remarks on time-space
   noncommutative field theories}, JHEP {\bf 0105} (2001) 057.\\
%
   A. Bassetto, L. Griguolo, G. Nardelli and F. Vian: {\it On the
   unitarity of quantum gauge theories on noncommutative spaces},
   JHEP {\bf 0107} (2001) 008.\\
%
   C-S. Chu, J. Lukierski and W.J. Zakrzewski, {\it Hermitian
   analyticity, IR / UV mixing and unitarity of noncommutative field
   theories}, {\tt hep-th/0201144}.  
%
%
\bibitem{Campbell}
   B.A. Campbell and K. Kaminsky, {\it Non-commutative field theory
   and spontaneous symmetry breaking}, Nucl. Phys. {\bf B581} (2000)
   240; {\it Non-commutative linear sigma models}, Nucl. Phys. {\bf 
   B606} (2001) 613.
%
\bibitem{Sarkar} 
    S. Sarkar and B. Sathiapalan, {\it Comments on the
    renormalizability of the broken symmetry phase in noncommutative
    scalar field theory}, JHEP {\bf 0105} (2001) 049.

\bibitem{Petriello}
   F. Petriello, {\it The Higgs mechanism in noncommutative gauge
   theories}, Nucl. Phys. {\bf B601} (2001) 169.
%
\bibitem{Liao}
   Y. Liao, {\it One-loop renormalization of spontaneously broken
   $U(2)$ gauge theory on noncommutative spacetime}, JHEP {\bf 0111}
   (2001) 067; {\it One-loop renormalizability of spontaneously broken
   gauge theory with a product of gauge groups on noncommutative
   spacetime: the $U(1)\times U(1)$ case}, {\tt hep-th/0201135}. 
\end{thebibliography}
\end{document}